%% file: MSNPDiscovery.tex
\documentclass{IOS-Book-Article}

\usepackage{mathptmx}

\usepackage{setspace}
\usepackage{cite}
\usepackage{url}
\usepackage{dsfont}
\usepackage{enumerate}
\usepackage{algorithm}
\usepackage{algorithmic}
\usepackage{amsmath}

\usepackage{subfig}
\usepackage{wrapfig}
\usepackage{array}
\usepackage{graphicx}
\usepackage{epstopdf}

\usepackage{array,ragged2e}

\usepackage{courier}

\usepackage{color}
\definecolor{dkgreen}{rgb}{0,0.6,0}
\usepackage{xcolor}
\usepackage{listings}
\usepackage{pbox}
\usepackage{caption}

\begin{document}
\begin{frontmatter}              

\title{Service Discovery and Trust in\\Mobile Social Network in Proximity}
\runningtitle{IOS Press Style Sample}

\author[A]{\fnms{Chii} \snm{Chang}%
\thanks{E-mail:
chiichang1@gmail.com}} 
and
\author[B]{\fnms{Satish} \snm{Srirama}%
\thanks{E-mail:
srirama@ut.ee}} 
and
\author[A]{\fnms{Sea} \snm{Ling}%
\thanks{E-mail:
chris.ling@monash.edu}} 


\runningauthor{B.P. Manager et al.}
\address[A]{Faculty of Information Technology, Monash University, Australia}
\address[B]{Institute of Computer Science, University of Tartu, Estonia}

\begin{abstract}
Service-oriented Mobile Social Network in Proximity (MSNP) lets participants establish new social interactions with strangers in public proximity using heterogeneous platforms and devices. Such characteristic faces challenges in discovery latency and trustworthiness. In a public service-oriented MSNP environment, which consists of a large number of participants, a content requester who searches for a particular service provided by other MSNP participants will need to retrieve and process a large number of Service Description Metadata (SDM) files, associated semantic metadata files and identifying the trustworthiness of the content providers. Performing such tasks on a resource constraint mobile device can be time consuming, and the overall discovery performance will be affected and will result in high latency. This paper analyses the service discovery models of MSNP and presents corresponding solutions to improve the service discovery performance of MSNP. We firstly present and analyse the basic service discovery models of service-oriented MSNP. To follow up, we apply a context-aware user preference prediction scheme to enhance the speed of the semantic service discovery process. Later, we address the trustworthiness issue in MSNP and propose a scheme to reduce the latency of the trustworthy service discovery for MSNP. The proposed scheme has been tested and evaluated on MSNP application prototype operating on real mobile devices and MSNP simulation environments.
\end{abstract}

\end{frontmatter}

\thispagestyle{empty}
\pagestyle{empty}

\input{Introduction}	
\input{RelatedWorks}
\input{Overview}

\input{Discovery_v2}
\input{Trust}
\input{Evaluation}
\section{Summary and Future Research Direction}
In this paper, we have presented the context-aware proactive service discovery scheme for service-oriented MSNP to reduce service discovery latency. Apart from service discovery in MSNP, trustworthiness has also been addressed. While existing work in trust management of MP2P environment focused on the trust model, and did not consider data transmission overhead issues, we have presented a lightweight trustworthy service discovery scheme specifically for service-oriented MSNP.

Although the proposed schemes in this paper can enhance the overall service discovery performance, the dynamic nature of MSNP environment can still lead to unpredictable situations in which mobile devices cannot perform the service discovery effectively. For example, a service advertiser attempting to advertise its service by utilising push-based approach may suddenly run low on hardware resource availability due to the sudden increase in the number of MSNP participants joining the environment. Another example, during trustworthy service discovery, a requester may only have the option to refer to the reputation rating from public proximal recommenders in which the requester has to identify the credibility of each candidate recommender. If the number of candidate recommenders is large, the overall process of determining trust can create serious latency. There is therefore a need to utilise the task offloading mechanism of mobile cloud computing dynamically at runtime when the mobile device is unable to handle an unexpected situation in service-oriented MSNP. 

We have identified several subjects for future research directions.
\begin{description}
\item \textbf{\textit{Social-aware service discovery for MSNP}}---Our context-aware user preference associated proactive service discovery for MSNP scheme requires a fair amount of context information associated service discovery and interaction records in order to predict the user preferred service type. However, for a user who has none or  a few records, the scheme is unable to predict the user preferred service type regarding to the context information of the user's current environment. One possible solution is to utilise the context associated service interaction records from the user's social groups such as friends or friends of a friend based on some similarity measurement between the user and his/her friend's profiles. However, a proactive service discovery scheme that relies on social-driven information will incur additional data retrieval cost at runtime, which can increase the overall service discovery makespan. A proper solution to overcome such an issue requires further investigation.
\item \textbf{\textit{A lightweight trustworthy service discovery for public MSNP}}---The proposed scheme for lightweight trustworthy service discovery is applicable when the user has a fair amount of reputation rating data from friends or friend of a friend. If the reputation rating data comes from the public, it is inevitable that the requester has to retrieve all the available reputation rating data from the available proximal MSNP participants. The transaction cost can also be high when the environment consists of a large number of proximal MSNP participants. 
\end{description}








\section*{Acknowledgement}
This research is supported by the European Regional Development Fund through the EXCS, Estonian Science Foundation grant PUT360.

\bibliographystyle{IEEEtran} 
\bibliography{IEEEabrv,MSNPDiscovery}
\end{document}

%% file: Introduction.tex
\section{Introduction}
\subsection{Preamble}
Accessing Social Network Services (SNS) such as Facebook\footnote{See \url{https://www.facebook.com/}}, Twitter\footnote{See \url{https://twitter.com/‎}} or Google+\footnote{See \url{https://plus.google.com/}} has become a common daily activity for Internet users. The marketing report \cite{comScore2012} shows that on average, a Facebook user accesses Facebook's services for over 7 hours per month, and with around 80\% usage traffic derived from mobile applications. 
Recent smart mobile devices, such as smartphones, tablets, handheld media players are capable of letting users produce various digital content, and share/upload the content to many SNS directly via wireless network connections. This has increased the number of people using mobile SNS applications. 

Although the marketing report has shown that mobile SNS applications have successfully become the most popular applications for mobile users, mobile users have been restricted in the virtual communities of online SNS and are not aware of the social opportunities available to them. While mobile users spend most of their time accessing the online SNS, they have missed many opportunities to interact with others for new friendships, business opportunities, or information sharing \cite{Borcea2007}. Consequently, applications such as those by  MobiSoC \cite{Borcea2007}, MobiClique \cite{Pietilainen2009}, MoSoSo \cite{Tsai2009}, Uttering \cite{Allen2010} and Spiderweb \cite{Sapuppo2010} have been proposed to enable a new breed of Mobile Social Network (MSN) functions, which can assist mobile users to interact with proximal people and perform various social activities such as searching for new friends who have common interests, exchanging content of common interests and establishing conversations. In this paper, such a proximal-based MSN environment is termed a \textbf{M}obile \textbf{S}ocial \textbf{N}etwork in \textbf{P}roximity (MSNP)  \cite{Chang2012ICSOC,Chang2013PMC,Chang2013JNIT,ChiiChangPhDThesis}. 

An MSNP should not be seen as a replacement of existing SNS but as its complement \cite{Pietilainen2009}. MSNP leverages online SNS with a proximal mobile wireless network connection by providing location-based social networking opportunities. It can be applied in various social scenarios. For example, with the assistance of MSNP applications, an attendee in a highly populated conference can easily find someone who has common interests based on information derived from public profiles and their public information on online SNS. Another example: visitors who attend a big exhibition such as Comiket\footnote{See \url{http://www.comiket.co.jp/index_e.html}} in Japan or Comic-Con in USA\footnote{See \url{http://www.comic-con.org/}} and Australia,\footnote{See \url{http://www.ozcomiccon.com/press.aspx}}  may be at a loss as to where they can find something they are interested in. With MSNP applications, they can rapidly discover the information about any point-of-interest shared by other MSNP application users based on their preferences in the same exhibition. MSNP also provides opportunities for active MSNP application users to bring more visitors to their online SNS spaces. For example, Twitter users can actively advertise content to MSNP application users who are potentially interested in the content, in order to bring more followers to their Twitter.

Although several software frameworks have been proposed to enable MSNP, most of these works are tightly coupled systems. In the past decade, service-oriented architecture (SOA) has become the mainstay of networked application development. Within SOA developments, the standardised Web service technologies\footnote{See \url{http://www.w3.org/2002/ws/}} that provide  platform independent common intercommunication interfaces have been broadly applied in various networked distributed computing areas to enhance the interoperability of different machines with heterogeneous software platforms. Web service has also been applied to numerous mobile applications either by utilising mobile applications as Web service clients \cite{Kang2007,Tian2007,Yoshikawa2003,Zahreddine2005}, or embedding HTTP Web servers to provide Mobile Web Service (MWS) directly from mobile devices \cite{Gehlen2005,Srirama2006,Koskela2007,Pawar2008a}. Utilising Web services can enable loose coupling and platform independent features for MSNP environments. 

Our research focuses on realising a loosely coupled, service-oriented MSNP environment based on Web service standard technologies. Service-oriented MSNP provides an open standard environment. With open standards, mobile application developers can easily implement compatible applications for mobile users to participate in MSNP without being bound to a particular device or software platform.

\subsection{Research Motivations}
Imagine an MSNP environment with a verity of mobile device users. Each user intends to use their MSNP application to interact with one another. However, due to each MSNP application being implemented in different technologies, the opportunity of discovery and interaction between mobile users is much less. For example, a user who is using an MSNP application  based on JXTA (Juxtapose)\footnote{See \url{http://java.sun.com/othertech/jxta/index.jsp}} will be unable to communicate with a user who is using the Universal Plug and Play (UPnP)-based\footnote{See \url{http://www.upnp.org/}} MSNP application, because the way they perform discovery is different. Moreover, when the environment grows, the number of operation types and content types increases. In order to fulfil the need, semantic annotation may be applied to describe the operations provided by each participating device. However, due to the heterogeneity issue, the semantic discovery mechanism is difficult to be implemented. Conversely, if the entire system is Web service compliant, the overall interoperability can be highly improved. 

The benefit of applying Web service standards in MSNP is explicit. However, enabling decentralised Web service-oriented MSNP faces a number of challenges:
\begin{itemize}
\item \textbf{Service Discovery Latency}---In service-oriented MSNP, each user's mobile device is a Web service client and also a Web service provider  \cite{Srirama2006}. Since Web service has been applied as the common communication interface, and in most cases, MSNP participants do not have pre-knowledge about other peers in order to support the service discovery process, each MSNP peer can use semantic Web standards, such as Web Services Description Language (WSDL), Extensible Markup Language (XML), Semantic Annotation for WSDL and XML Schema (SAWSDL)\footnote{See \url{http://www.w3.org/TR/sawsdl/}} and Web Ontology Language (OWL)\footnote{See \url{http://www.w3.org/TR/owl2-overview/}} to describe its services. While performing service discovery, an MSNP peer has to retrieve and process the other participants' service description metadata at runtime to enable dynamic Web service binding. Moreover, an MSNP peer is required to perform trust control processes on mobile devices in the mobile peer-to-peer (MP2P) network environment. Such a discovery process can cause high latency when the environment consists of large number of mobile Web service providers. Moreover, the dynamic nature of MP2P requires the service discovery process to be fast in order to enable further interaction processes, because MSNP peers are extremely mobile. Each can move out from the current Wi-Fi network and join another, or can switch between 3G/4G mobile Internet. 

\item \textbf{Trust}---Suppose a content requester discovers a Web service from a previous unknown content provider who can provide content of interest to the requester. Should the requester use the service to retrieve the content? In a basic process, the decision can be made by the human manually. However, if a more advanced autonomous service discovery operation is required, the task becomes critical. Imagine an MSNP participant intends to mashup a particular content from all content providers who provide the corresponding services. The process becomes inefficient if it requires the human user to manually select which providers' services they want to use. Hence, an autonomous decision making mechanism is more efficient, but trust is a major concern. Fundamentally, supporting trust in a Web service environment such as applying WS-Trust\footnote{See \url{http://docs.oasis-open.org/ws-sx/ws-trust/v1.4/ws-trust.html}} requires a global entity to manage the trust-related data. However, because service discovery in MSNP is based on MP2P topology, it is impossible to establish a global central management party for supporting trustworthiness \cite{Qureshi2010}. Hence, each MSNP participant has to manage the trust by itself. In a common SNS such as Facebook, a user can define different levels of content accessibility according to different social groups. A similar approach can be applied in a MSN solution \cite{Kourtellis2010}. However, for a new MSNP participant, who does not have many contacts, it is hard to define such access control.
\end{itemize}

This paper presents an extension of our previous works \cite{Chang2011,Chang2012ICSOC,Chang2013PMC,Chang2013JNIT} and provides the details in service discovery of MSNP. In this paper, we analyse the service discovery models of MSNP and presents corresponding solutions to improve the service discovery performance of MSNP. The two major challenges: Service Discovery Latency and Trust are analysed. Their corresponding solutions: a context-aware user preference-associated proactive service discovery scheme and a lightweight trustworthy service discovery scheme are proposed and evaluated.

This paper is organised as follow: Section 2 proivdes a literature review of related works. Section 3 provides an overview of service-oriented MSNP architecture. Section 4 presents and analyses the basic service discovery models of service-oriented MSNP. In Section 5, we apply a context-aware user preference prediction scheme to enhance the semantic service discovery process. In Section 6, we address the trustworthiness issue in MSNP and propose a scheme to reduce the latency of the trustworthy service discovery for MSNP. Section 7 presents the evaluation of the proposed schemes. Section 8 summarises our work and provides the future research directions in this research domain.

%% file: RelatedWorks.tex
\section{Background}
An ideal MSNP framework should support the following capabilities:
\begin{itemize}
\item \textit{Decentralised}, which can avoid single point of failure issues.

\item \textit{Autonomous discovery}, to support a mechanism to improve the discovery result.
\item \textit{Trust}, to support trustworthiness to help users interact with people who are not in their contact list.
\item \textit{Loose coupling}. To enhance the interoperability of a heterogeneous platform.
\item \textit{Latency reduction}, to provide a proper strategy to reduce the latency of the message-driven discovery, because a loosely coupled MSNP system faces latency challenges.
\end{itemize}

\subsection{Related Frameworks}
Existing related frameworks are still in their early stages. None of the frameworks described below supports all the above ideals of MSNP. 

A purely centralised framework such as MobiSoC \cite{Borcea2007} and MobilisGroups \cite{Lubke2011}  potentially harbours the risk of single-point-of-failure. Some centralised solutions such as Smart Campus Project \cite{Yu2011} and SPN \cite{Yang2008} support minor decentralised communication capabilities by utilising Bluetooth technology when the central server is not available. However, such a solution is insufficient, because by simply utilising Bluetooth-based discovery, it can result in high latency especially when the environment grows.

Most existing works also lack support for heterogeneous platform interoperability. They were proposed in the form of stand-alone technology. Within these frameworks, some have applied standard service-oriented technologies. MobiSoC is a Web service-based framework that applied SOAP communication. MobilisGroups has utilised IETF XMPP \cite{XMPPFoundation}, which is a popular centralised standard communication protocol. SocioNet is also a Web service-based framework. Yarta has utilised standard protocol---Service Location Protocol (SLP)\footnote{See \url{http://www.rfc-editor.org/rfc/rfc2165.txt}}---for proximal mobile P2P discovery. 

Within these related frameworks, Yarta is the closest framework to achieve the basic capabilities described previously. It is capable of avoiding a single point of failure, and it supports heterogeneous platform interoperability and autonomous discovery. However, Yarta has not provided a strategy to reduce latency caused by applying standard semantic discovery technology in a MP2P network. The evaluation result of Yarta's prototype has indicated that this is an issue. Further, Yarta has no support for resource-awareness, in which the discovery and interaction scheme should adapt to the resource changes and environmental factors. 

Overall, existing works did not address trustworthiness, which is an important aspect of MSNP because MSNP allows users to interact with new people who are not in their existing contact list. Without a proper strategy for trust, people will hesitate to use MSNP. Further, applying trust in MSNP can also cause additional latency in the \textit{bootstrap and discovery phase} because MSNP is based on MP2P topology in which the involved data for performing trust control is distributed (e.g., stored in each MSNP participants' backend cloud storage) and require the mobile application to retrieve them at runtime via the unstable mobile Internet. Hence, reducing latency for discovery phase becomes a priority challenge which needs to be resolved in MSNP. In the next section, we review a number of trustworthy service discovery solutions designed for MP2P environments.

\subsection{Trustworthiness in Mobile Peer-to-Peer Environments}
A number of works have been proposed to support trustworthiness in MP2P environments. While works proposed by Li et al. \cite{Li2010} and Rathnayake et al. \cite{Rathnayake2011} were focusing on how to improve the reliability of trust models by utilising the computation of a large number of trust-related data, resulting in  insufficient processing speed in MP2P network \cite{Niu2013}, some authors \cite{Wu2009,Qureshi2012,Waluyo2012} have proposed lightweight trustworthy service/peer discovery schemes for MP2P environments. 

Reducing data transaction is a common strategy to improve the processing speed of trust in MP2P. Wu et al. \cite{Wu2009} have proposed a group-based reputation scheme. Their design is based on super peer MP2P network, in which a super-peer (which is described as Power peer in their work) manages the reputation rating data of a group of mobile peers with similar movement speed. However, in a public environment such as MSNP, users may not be willing to let their devices act as super-peers because the high frequency of data transaction through their mobile devices can consume too much hardware resources (i.e. CPU, RAM, battery life, etc.). 

M-Trust \cite{Qureshi2012} reduces reputation data transaction by selecting recommenders based on the confidence of the candidate recommenders. A disadvantage of M-Trust is that the system will directly remove a trustworthy peer's recommendation (reputation rating) when the peer is disconnected from the current network (either due to network switching or due to the \textit{time to live} of its recommendation expiring). It would be ideal to provide a strategy to let M-Trust retrieve updates from recommenders in a different network, but this has not been addressed.

Similar to the fundamental strategy of M-Trust, TEMPR \cite{Waluyo2012} also improves the trust processing speed by utilising the selective recommender approach. Distinguished from M-Trust, the TEMPR scheme computes direct peers' (candidate recommenders who can directly interact with the requester) trustworthiness based on two scores: (1) the direct peers' trustworthy rating from other unknown peers; and (2) the direct peers' untrustworthy rating from other unknown peers. 

Our work in trustworthy service discovery can be seen as an extension of TEMPR, designed specifically for service-oriented MSNP. The major difference is that we do not assume strangers' application will always forward messages to assist other participants for the trust processes. Hence, a requester who intends to identify a provider's trustworthiness has to obtain the reputation rating data by either directly retrieving the reputation data from the other MSNP participants or by retrieving the data from the other MSNP participants' cloud storages.

%% file: Overview.tex
\section{Service-Oriented MSNP Architecture Overview}

MSNP represents an environment in which mobile users utilise their mobile devices to perform social activities with each other in proximal distance. The fundamental aim of MSNP is to enable communication in a fairly close range so that the participants can potentially meet each other. Figure 1 illustrates an MSNP environment. In order to improve the interoperability, Web service has been utilised as the common communication interface. 
\begin{figure*}[h]
\centering
  \includegraphics[width=0.9\textwidth]{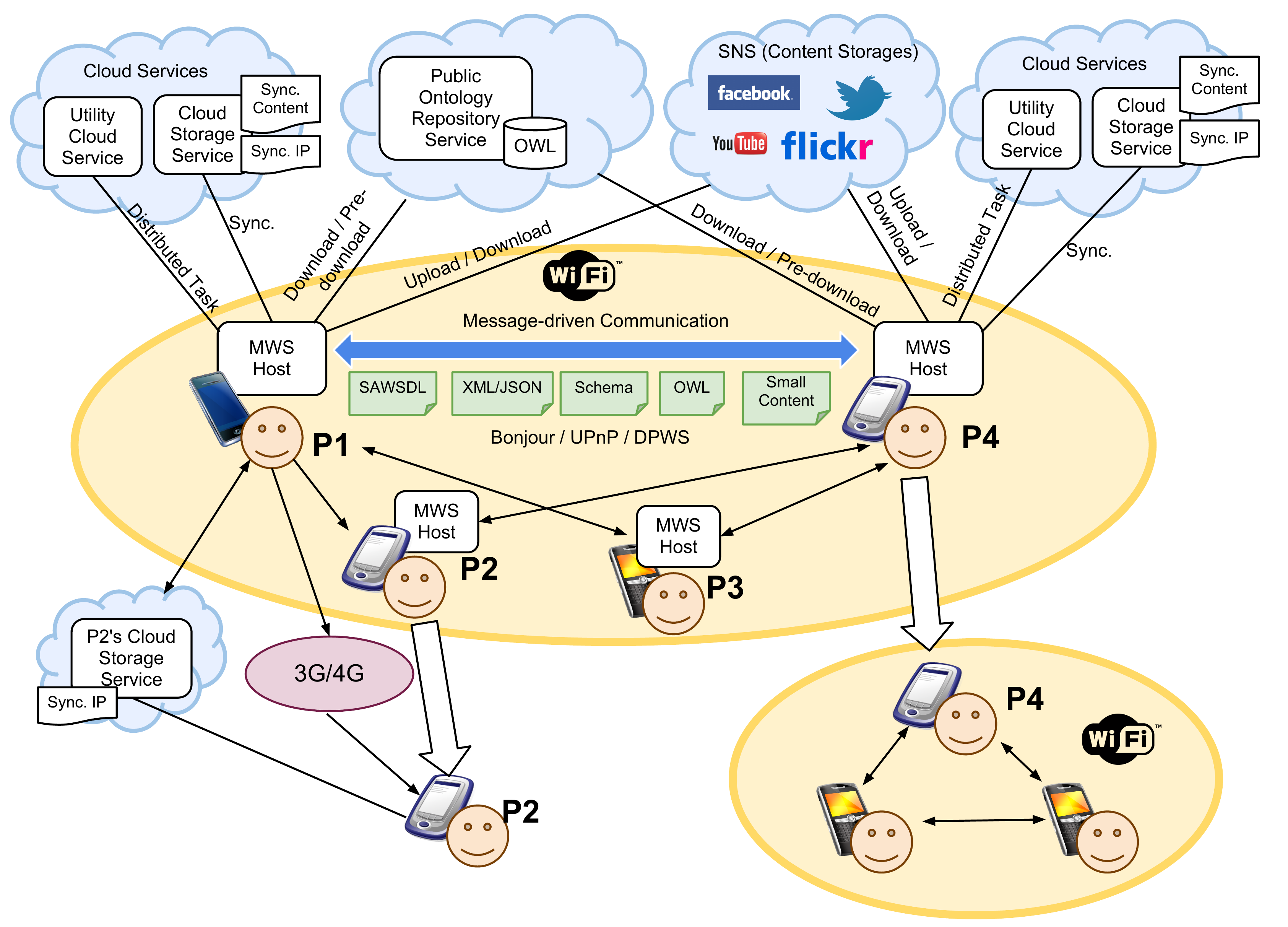}
  \caption{Service-Oriented MSNP Architecture.}
\end{figure*}
In an MSNP environment, each mobile device is a mobile Web service consumer and also a provider \cite{Srirama2006}. When two peers join the same wireless network, they utilise standard communication technologies such as DPWS \cite{OASIS2009} or Zeroconf \cite{IETF} to exchange their Service Description Metadata (SDM). For peers who do not have Mobile IPv6,\footnote{See \url{http://tools.ietf.org/html/rfc6275}} we expect each to have its own back-end cloud storage to synchronise its IP address as a small text file in its cloud storage (or alternatively utilising public Domain Name System [DNS] servers if available). The Uniform Resource Locator (URL) of the text file is described in a peer's SDM. Hence, when the peer (e.g., Figure 1, P2 or P4) moves out from the current network, the other peers (e.g., Figure 1, P1 and P3) in their previous network can still interact with P2 or P4 via mobile Internet.

Since P1 and P3 have previously exchanged their SDM with P2 and P4, they have cached the SDM of P2 and P4 in either their local memory or synchronised it to their cloud storages. When P1 and P3 receive requests from other peers in the same network that are performing service discovery, P1 and P3 can also provide P2 and P4's SDM to these requesting peers. Instead of having the SDM directly sent to the peers by P1 and P3, P1 and P3 can synchronise the cached SDM to their cloud storages, and simply provide the URL link to the requesting peers. 

A similar concept can be applied to content sharing and mashup, say for example, P1 intends to mashup the content provided by P2 and P3. When P1 invokes P2 and P3 for the content, P2 and P3 will simply reply with the corresponding metadata documents, which contain the description about where the content can be retrieved from in the Internet. For example, P2 has uploaded the content to a SNS as public accessible content. Hence, P2's response metadata will contain the URL link of the uploaded content.

Taking into account that mobile devices usually have limited processing power, it is reasonable for an MSNP peer to delegate some of its processes to its backend Cloud Utility Service (CloudUtil). In Figure 1, for example, P1 utilises its backend CloudUtil for semantic service discovery. Further, CloudUtil can also be used to directly access the content uploaded by other MSNP peers in SNS to discover useful content for P1's mashup (if the content has been described in Rich Site Summary [RSS]\footnote{See \url{http://tools.ietf.org/id/draft-nottingham-rss-media-type-00.txt}} feed format).

A content provider in MSNP can also actively push recommendations to other participants based on the participants' service preferences. Due to privacy concerns, MSNP peers may prefer not to share their private information. However, when a list of available services (described semantically) is provided to the participants, the participants can simply reply which service type they are interested in.

%% file: Discovery_v2.tex
\section{Service Discovery in Service-Oriented MSNP}

This section presents and analyses the basic service discovery models of service-oriented
MSNP. Before we proceed with our discussion, we define and reiterate the terminologies in a service-oriented MSNP:
\begin{itemize}
\item A \textit{device} represents a mobile device such as a smart phone, a handheld media player or a small tablet computer. A device can be operated by any operating system and is capable of participating in mobile P2P network using software applications.
\item An \textit{agent} represents an MWS-enabled software agent. The term---agent is derived from the software agent described in W3C Web Service Architecture document \cite{website:W3CWebServiceArchitecture}, in which an agent performs Web service activities for its human user. In MSNP, an agent can perform functions for both MWS client and server.  
\item A \textit{user} is a human user who holds a mobile device that is embedded with mobile Web service (MWS)-enabled software agent.
\item An \textit{MSNP participant} represents an entity which participates in an MSNP environment. Each MSNP participant consists of: a \textit{device}, an \textit{agent}, and a \textit{user}.
\end{itemize}

In MSNP, each agent would have pre-downloaded a fair number of public common ontologies that have been published on cloud resources (e.g., Swoogle\footnote{\url{http://swoogle.umbc.edu/}}, or FUSION\footnote{\url{http://www.seerc.org/fusion/semanticregistry/}}). A public common ontology describes numerous common service types and data types semantically. Each Semantic Annotations for WSDL and XML Schema (SAWSDL) \cite{Akkiraju2007}-compliant agent describes its services using semantic annotations that map to the corresponding ontology types. Benefiting from the public common ontologies and semantic annotation, an MSNP service requester's agent can identify whether a service matches to the functionality it needs from the service provider's WSDL and related documents (e.g., XML Schema). In the following subsections, we discuss service discovery models in MSNP. 

\subsection{Pull-based Service Discovery}

Pull-based service discovery  in service-oriented MSNP represents the most basic service discovery mechanism that is supported by the existing mobile P2P protocols (e.g., UPnP, Bonjour, DPWS, etc.) without making a significant assumption, such as expecting the requester \textit{agent} to provide MWS to let other \textit{agents} advertise SDM to it.

Figure 2 illustrates the process flow of the pull-based service discovery model described in Business Process Modelling Notation (BPMN)\footnote{\url{http://www.bpmn.org/}}.
\begin{figure*}[h]
  \centering
    \includegraphics[width=0.75\textwidth]{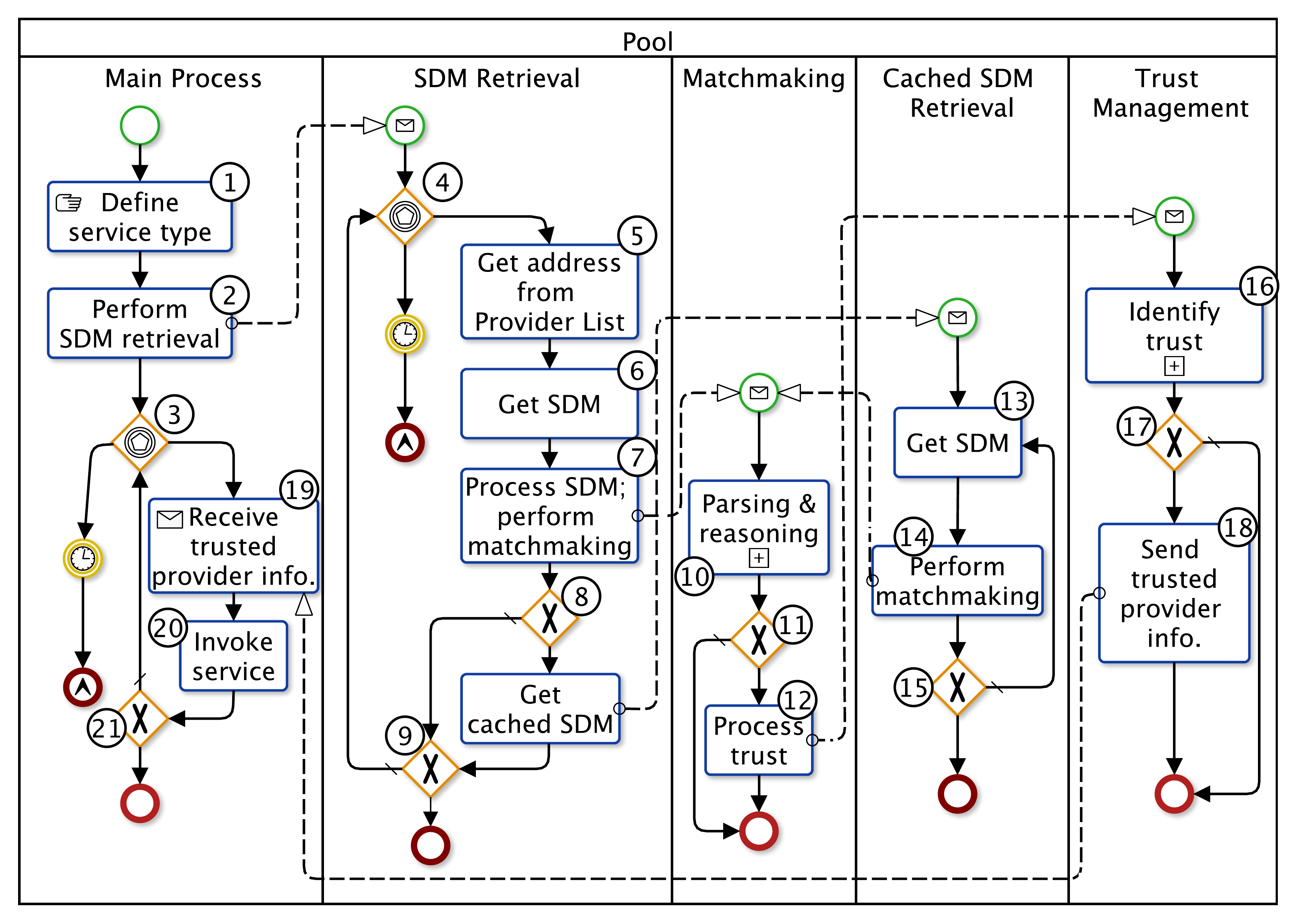}
    \caption{Pull-based service discovery in MSNP}
\end{figure*}
The discovery process consists of five subprocesses:
\begin{itemize}
\item \textbf{Main Process}. When a user joins an MSNP environment, he/she manually defines his/her need (task 1) and then requests his/her $agent$ (denoted by $agent_{rqt}$ to discovery the corresponding service provided by other MSNP participants' $agents$ (task 2). The $agent_{rqt}$ launches the SDM retrieval subprocess and keeps the main process thread on stand by waiting for the result from the Trust Management subprocess (task 19). When a trusted provider information is passed to the main process, $agent_{rqt}$ will invoke the corresponding service from the provider $agent$ to retrieve the result (task 20).
\item \textbf{SDM Retrieval}. The SDM Retrieval subprocess is set to a finite timestamp (mark 4). When time is up, this subprocess will be terminated. The main activity of this subprocess is to retrieve SDM from each MSNP participant's $agent$ in the requester's current environment (task 5, 6). After retrieving and processing the SDM, $agent_{rqt}$ may find out that the provider $agent$ is also providing a service which returns a  list of other $agents$' SDMs which were fetched when the provider $agent$ performed discovery previously when it joined the current environment. If such a \textit{cached SDM} service is available, $agent_{rqt}$ will launch the \textbf{Cached SDM Retrieval} subprocess.
\item \textbf{Cached SDM Retrieval} subprocess retrieves one or more cached SDMs from the provider (task 13). The cached SDMs can either be retrieved from its provider $agent$ directly or be retrieved from the provider's cloud storage depending on the provider's preference. The retrieved SDM is also passed to the Matchmaking subprocess (task 14).
\item \textbf{Matchmaking}. The retrieved SDM and its associated documents will be processed by the Matchmaking subprocess. The Matchmaking subprocess uses semantic reasoning algorithm and XML document parsing technologies to identify whether the provider can provide a corresponding service to fulfil the request or not (task 10). If the provider can provide the corresponding service, $agent_{rqt}$ will perform \textbf{Trust Management} (task 12).
\item \textbf{Trust Management} subprocess identifies whether the provider's service is trustworthy or not (task 16).  If the provider's service is trustworthy, $agent_{rqt}$ will perform the service invocation to retrieve result from the provider (task 18).
\end{itemize}

A drawback of this simple model is the latency issue. Because SDMs are described in XML format, resource-constraint mobile devices are usually unable to process a large number of XML documents effectively. 

\subsection{Push-based Service Discovery}
Push-based service discovery approach involves a requester \textit{agent} ($agent_{rqt}$) utilising passive mechanism to receive SDMs advertised by the other active MSNP participants' $agents$. Figure 3 illustrates the process flow of push-based service discovery approach. 
\begin{figure*}[!h]
  \centering
    \includegraphics[width=0.75\textwidth]{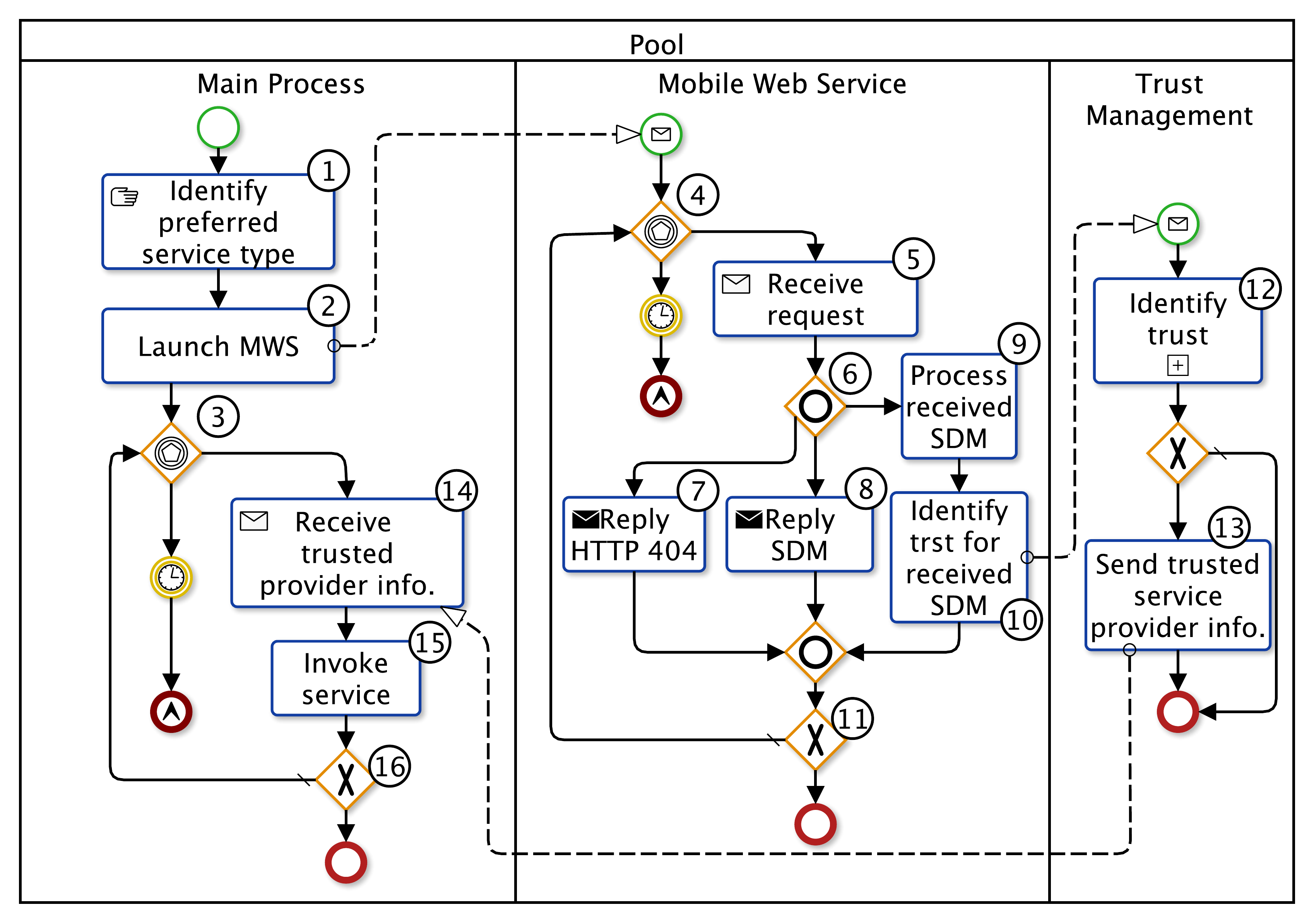}
    \caption{Push-based service discovery in MSNP}
\end{figure*}

The behaviour of the main components in this approach is described below:
\begin{itemize}
\item \textbf{Main Process} in push-based approach differentiates from the pull-based approach in that $agent_{rqt}$ does not actively invoke the other participants' $agents$ to retrieve their SDM. Instead,  $agent_{rqt}$ launches an MWS provider (task 2) to passively receive SDM advertised by the other $agents$.
\item \textbf{Mobile Web Service} subprocess permits the requester to be passive. In this subprocess, $agent_{rqt}$ provides MWS to let other participants' $agent$ retrieve $agent_{rqt}$'s SDM (task 8). Based on $agent_{rqt}$'s SDM, other $agents$ directly push their SDM to $agent_{rqt}$. When $agent_{rqt}$ receives an SDM, it performs the matchmaking task to identify whether the SDM's provider can provide the required service type or not (task 9). If the SDM's provider can provide the required service type, $agent_{rqt}$ will perform the Trust Management subprocess to identify the provider's trustworthiness (task 10). 
\item \textbf{Trust Management} subprocess is the same as the Trust Management process described in the previous pull-based model. 
\end{itemize}

\subsection{User Preference Associated Push-based Service Discovery}
In addition to the two basic service discovery approaches described in the previous sections, we propose a new service discovery approach for MSNP---the user preference associated push-based service discovery (\textit{PrefPush}). The relative works of \textit{PrefPush} have been previously published in \cite{Chang2011,Chang2013JNIT,Chang2013PMC}.  \textit{PrefPush} in MSNP relies on participants' $agents$ actively advertising their SDM to one another. A requester participant's agent---$agent_{rqt}$ will perform the following three subprocesses to enable \textit{PrefPush}. Figure 4 describes the process flow of the \textit{PrefPush}-based service discovery model in BPMN.
\begin{figure*}[!h]
  \centering
    \includegraphics[width=0.75\textwidth]{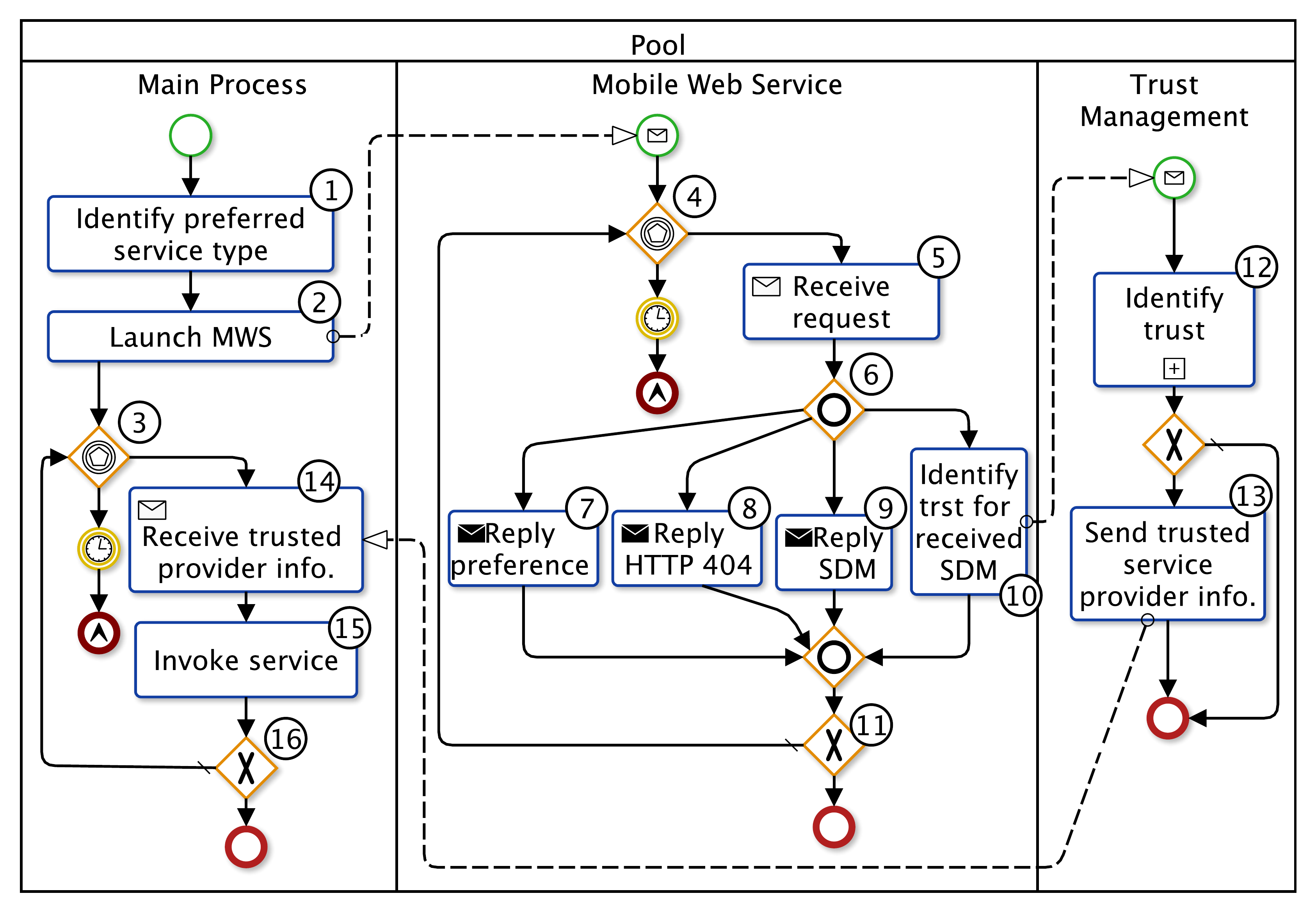}
    \caption{\textit{PrefPush}-based service discovery in MSNP}
\end{figure*}
\begin{itemize}
\item \textbf{Main Process} is slightly different from the pull-based model. In the \textit{PrefPush} approach, when a user joins the environment, $agent_{rqt}$ can autonomously identify its user's preferred service in the current environment based on the user's past request records and the current environmental context information (task 1). The details of the approach to enable the autonomous identification of the user's preferred service will be described in the next section. Alternatively, user can manually define his/her preferred service type prior or on demand at runtime. Once the preferred service type is identified, $agent_{rqt}$ launches its Mobile Web Service (MWS) server-side mechanism to let other participant's $agents$ actively interact with it (task 2). Afterwards, $agent_{rqt}$ puts the main process on stand by waiting for the result from the Trust Management subprocess (task 14). When a trusted provider information is returned from the Trust Management subprocess, $agent_{rqt}$ will invoke the trusted provider's service to retrieve the result.
\item \textbf{Mobile Web Service} subprocess enables the requester to be passive. In this subprocess, $agent_{rqt}$ provides MWS to let other participants' $agents$ retrieve $agent_{rqt}$'s SDM (task 9). Based on the $agent_{rqt}$'s SDM, other participants' $agents$ can also request $agent_{rqt}$ for its user preferred service type (mark 7). The other participants' $agents$ who retrieved $agent_{rqt}$'s user preferred service type information, can determine whether their services can fulfil the need or not. If they can, they can post their SDM to $agent_{rqt}$ as advertisement. Note that, at this stage, we do not expect the other participants' $agent$ to directly post the content corresponding to the $agent_{rqt}$'s user preferred service type, because $agent_{rqt}$ does not know whether the provider is trustworthy or not. Hence, $agent_{rqt}$ expects a SDM advertisement rather then the corresponding content. Once $agent_{rqt}$ receives a SDM, it performs the Trust Management subprocess to identify the trustworthiness of the provider (task 10).
\item \textbf{Trust Management} subprocess is the same as the Trust Management process described in the previous pull-based model.
\end{itemize}

The major difference between \textit{PrefPush} and the previous two approaches (\textit{Pull} and \textit{Push}) is that in the \textit{PrefPush}-based model, $agent_{rqt}$ does not need to perform any SDM retrieval process or perform semantic matchmaking process. Since the matchmaking process is done by other participants' \textit{agents}, the overall discovery makespan of \textit{PrefPush} can be much lower than the other two approaches. 

However, the drawback of \textit{PrefPush} is that it assumes  other participants' $agents$ remain active and will advertise their SDM to the others. Since such an assumption cannot be guaranteed, it is more feasible to perform both pull-based and \textit{PrefPush}-based model at the same time for the service discovery. 

\subsection{Hybrid-based Service Discovery}
Hybrid-based service discovery model in MSNP combines both pull and \textit{PrefPush}-based service discovery models. Figure 5 illustrates a hybrid-based service discovery model. The service discovery process consists of three main parallel tasks: pull-based service discovery, \emph{PrefPush}-based service discovery, and the trust management. 

\begin{figure}[!h]
  \centering
    \includegraphics[width=0.45\textwidth]{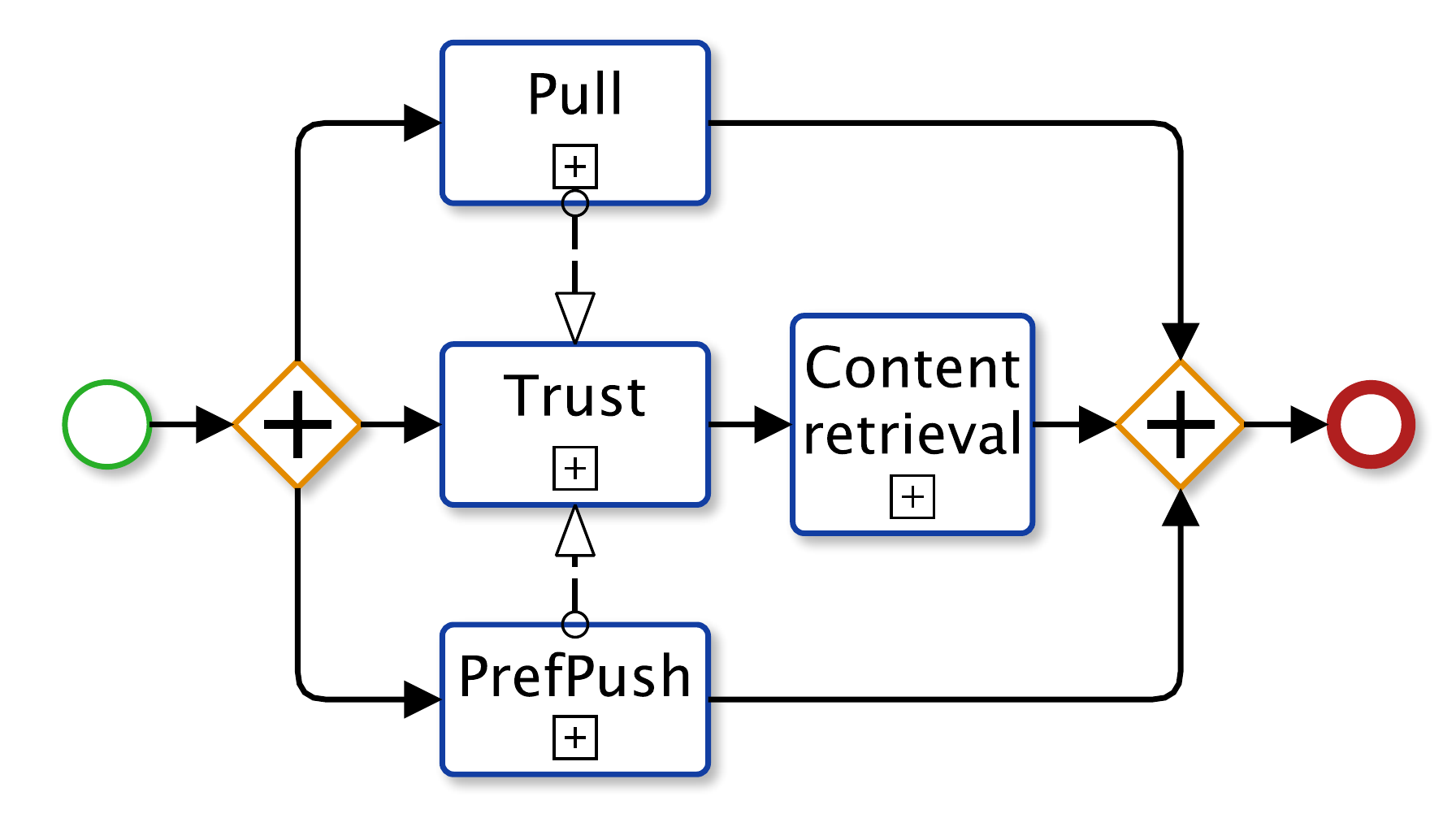}
    \caption{Hybrid-based service discovery in MSNP}
\end{figure}

When an MSNP user enters an MSNP-enabled environment, he can either manually launch the application to search for his/her preferred service provider, or his/her \textit{agent} can automatically triggers service discovery mechanism at background based on the user's preference computed from the scheme described in Section 5. The \textit{agent} will use a client-side MWS mechanism to search service providers and also launch an MWS server to allow other participants' \textit{agents} to actively advertise their SDM to it.

The pull-based service discovery task and \textit{PrefPush}-based service discovery task will pass SDMs retrieved from other service provider \textit{agents} to the trust management task. If the provider is trustworthy, the \textit{agent} will interact with the provider to retrieve the result/content as described in the previous two models.

\section{Context-Aware Proactive Service Discovery in MSNP}
Dey \cite{Dey2001} defined context as: \textit{`any information that can be used to characterise the situation of an entity. An entity is a person, place, or object that is considered relevant to the interaction between a user and an application, including the user and applications themselves.'} In the following paragraphs, we use Dey's definition as the basis to describe context in this paper.

Push-based service discovery in MSNP can be greatly improved by applying a proactive autonomous discovery mechanism. As Figure 3 shows, in the first task in the Main Process, if it relies on user manually entering the preferred service type, after the service type is entered, the user has to wait until the result returns. However, if the agent can predict the user preferred service type, it can support the \textit{PrefPush}-based service discovery approach in which the agent can autonomously start the discovery process when the user enters the MSNP environment. By doing so, when the user starts using the application, the agent has already discovered and identified a list of trusted services provided by the others. Moreover, some results may have already been pre-fetched by the agent if the user has granted the agent to do so. 

In this section, we present our proposed context-aware proactive service discovery scheme for MSNP. The proposed scheme enables the agent to perform SDM prefetching and content prefetching to reduce service discovery latency. 

Before the details of the proposed scheme are discussed, we provide the background of the proposed scheme.

\subsection{Background of Proactive Service Discovery}
The fundamental part of the user preference associated push-based service discovery approach proposed is based on the autonomous data prefetching mechanism. In general, the prefetching mechanism consists of the elements described in the following sections.

\subsubsection{Prediction}
The prediction mechanism aims to predict a user's request based on various factors. Factors in a prefetching approach for Web browsing \cite{Jiang1998,Tuah2003,Burklen2006} include user's browsing history, interests, navigation behaviour, and the popularity of the available contents/resources. By analysing these factors and comparing them with presently available contents, the probability of user's interest in a content can be computed. 

In mobile and pervasive computing environments, more factors need to be considered \cite{Drew2004,Choi2005,Feng2006,Drakatos2009,Jin2007,Boldrini2010}. These are user's current location, moving direction, hardware resources, network bandwidth, and many others.  Based on these factors, corresponding policies or rules can be designed and applied to a decision-making scheme to predict and anticipate a mobile user's future request more accurately. 

In Web-based systems, Jiang and Kleinrock \cite{Jiang1998} have introduced a prediction module to track user's access history continuously. Based on the historical records, the system can compute the probability of user's browsing actions, and determine what content needs to be prefetched. An extended approach proposed by Tuah et al. \cite{Tuah2003}  applied compound access graph to perform prediction based on the most recent browsing histories and the relationships between web pages. A mobile environment based prefetching scheme proposed by B\"{u}rklen et al.\cite{Burklen2006} has considered the location factor. The prediction result was calculated based on the user's searching histories in specific locations. These techniques were proposed for Web systems and their prediction decision modules only considered static factors. They did not consider dynamic factors such as the user's preference in different situations and events. 

A number of researchers \cite{Drew2004,Choi2005,Burklen2006,Drakatos2009} have proposed location-based and movement-based prediction scheme for cache prefetching. These works predict the probability of user's future query by analysing the user's present and future locations (based on his/her movement prediction), the corresponding query history records, and the predefined user preference profiles. However, in reality, a user's preference can dynamically be changed at runtime due to other factors. Moreover, the pre-defined static user preference profiles and rules are difficult to fulfil unanticipated situations \cite{Chen2005}, unless the user is willing to adequately define many different preferences manually for all possible situations. In most cases, a user is unable to define his or her probability for events accurately \cite{Heckerman1996}. Therefore, a proper adaptive scheme is required in the prediction mechanism.

\subsubsection{Adaptation and Context}
Adaptivity is an important concern in the autonomous data prefetching approaches, especially in the resource-constrained mobile computing environments in which network bandwidth, and hardware resources (i.e., cache size, energy) are limited. Without a properly designed strategy, a prefetching scheme may incur excessive resource costs \cite{Yin2002,Drakatos2009,Pallis2008}.   

A number of researchers have proposed approaches to improve the adaptivity of their prefetching schemes in different aspects. An earlier work proposed by Jiang and Kleinrock \cite{Jiang1998} was concerned with system resource usage. In their approach, the prefetching behaviour was dynamically adjusted based on the access performance. In the work proposed by Pallis et al. \cite{Pallis2008}, a policy and proxy-based prefetching strategy was proposed. Service consumers have been categorised into different cluster groups based on their interests, so the proxy can prefetch data more efficiently, and reduce the bandwidth cost. Yin et al. \cite{Yin2002} proposed a value-based adaptive prefetch (VAP) scheme, in which each data item has been assigned a value. Based on the assigned value, the current remaining power level, the access rate, the update rate, and the data size, the system can evaluate the cost of prefetching, and adjust the prefetching decision dynamically. Hu et al. \cite{Hu2003} proposed the Sliding Cache technique for adaptive prefetching, in which the cache space was dynamically changed based on the usage of the cached data item. Their evaluation showed that the approach can reduce the frequency of prefetching processes, and the results showed that the lesser the frequency of prefetching the lower the energy cost. 

Improving the accuracy of prefetching is one of the most important aspects to improve adaptivity. Drakatos et al. \cite{Drakatos2009} proposed a context-aware cache management prefetching strategy. The proposed cell-based mobility scheme is capable of detecting user's movement, and predicting use's future location. Based on the predicted future location together with query patterns (previous query records), the system is able to prefetch data item more accurately. The authors have also mentioned that if a user's preference model has been applied, the accuracy of prefetching can be explicitly improved. However, further detail in this respect was not elaborated in their works. 

User preference profiling is one of the major aspects to improve the accuracy of the prediction strategy. When accuracy is increased, the overall adaptivity is also improved due to the resource costs being reduced.  However, existing works \cite{Burklen2006,Choi2005,DelPrete2010} did not consider the dynamism of the user's preference. It is near impossible and inconvenient for most ordinary users to manually pre-define various preferences for all possible situations. The system needs to autonomously compute user's preference at runtime not only based on the historical query records, but also taking user's current context into consideration. To overcome this challenge, our proposed strategy aims to dynamically predict user preference at runtime using context-aware mechanisms.

\subsection{Context-aware User Preferred Service Prediction}
The main technique that ensures the success of proactive service discovery in our system is the context-aware prediction scheme. The context-aware prediction scheme takes user's current contexts as the basis, and then compares the current contexts to historical records to compute which query requested by the user has the highest probability. Each query recorded by the system has its associated \textit{semantic service type}. By predicting the highest probable query, the system is capable of identifying what \textit{semantic service type} is interested by the user in current environment. In this section, we describe our proposed context-aware prediction scheme. \newline

\noindent \textbf{Definition 1: \textit{Raw Context Data---$B$}.} $B=\{ b_i : 1 \leq i \leq N \}$. A $b_i$  is the data retrieved from context providers such as Global Positioning System, Compass application, image sensor, video sensor, voice sensor, and so on. A $b_i$ will be used as the basic input parameter to describe an interpreted context.\newline

\noindent \textbf{Definition 2: \textit{Interpreted Context---$C$}.} $C$ is a set of output from a rule-based context interpreting process, in which $C=\{ c_j : 1 \leq j \leq N \}$. Each $c_j \in C$ consists of $ID$, $type$, $value$, and a set of associated raw context data $B_{c_j}$. 

Based on Delir Haghighi et al.'s work \cite{DelirHaghighi2008}, an interpreting rule consists of context type ($type^{c_1}$), the scope of raw context data value, which includes minimum value and maximum value, and the output represents the interpreted value from this definition. For example, an interpreting rule describes \verb+inputMin="x12y14"+, \verb+inputMax="x37y22"+, \verb+type = "location"+, \verb+output ="MeetingRoom"+. When a retrieved location context contains a value: \verb+x15y17+, which is within the scope of \verb+inputMin+, and \verb+inputMax+, the system will consider the location \verb+"MeetingRoom"+ as one of the current contexts.\newline

\noindent \textbf{Definition 3: \textit{Query Records---$R$}.} Each device should maintain a set of query records $R$, in which $R=\{r_k : 1 \leq k \leq N\}$. $R$ represents the device user's previous queries associated with corresponding contexts. Each record $r_k$ consists of a query $q_{r_k}$ and a collection of context information $C_{r_k}$ occurred when $q_{r_k}$ is submitted by the user. 

A $q_{r_k}$ represents a request query submitted by the user for invoking either an internal embedded Web service on his/her device or an external Web service provided by other mobile device peers within the network. A $q_{r_k}$ consists of $ID$, $parameters$, and the corresponding \textit{semantic Web service operation type}. \newline

\noindent \textbf{Definition 4: \textit{Raw Candidate Queries---$Q$}.} $Q=\{q_l : 1 \leq l \leq N\}$. $Q$ is a set of non-duplicate queries from $R$:
\begin{equation}
Q=\bigcup_{k=1}^{ \left | R \right | }q_{r_k} 
\end{equation}

When the \textit{Predictor} component receives a set of contexts, it can predict the user's query based on the comparison result between the current contexts and the contexts of each query record. User may also define a preferred query manually by setting a set of context and the corresponding query in a file, which will be loaded in the beginning of the process. If user's definition exists, it will be used as the priority option. Otherwise, the system will perform the prediction automatically. 

Let $\tilde{C}$ be a set of current contexts, where $\tilde{C} = \{\tilde{c_i} : 1 \leq i \leq N\}$. By applying Bayes' theorem \cite{Clema1973}, the probability of a $q_l \in Q$ with one associated context $c_i$ can be computed from (2):
\begin{equation}
P(q_l | \tilde{c_i}) = \frac{P(\tilde{c_i}|q_l) \cdot P(q_l)}{P(\tilde{c_i})}
\end{equation}
where $P(\tilde{c_i}|q_l)$ is the probability of $\tilde{c_i}$ when $q_l$ was requested. It is computed from (3):
\begin{equation}
P(\tilde{c_i} | q_l) = \frac
{|\{r_k \in R : q_{r_k} \equiv q_l \wedge \exists c_x \in C_{r_k}, c_x \equiv \tilde{c_i} \}|}
{|\{ r_k \in R : q_{r_k} \equiv q_l \}|}
\end{equation}
$P(q_l)$ is theprobability number of occurrence of $q_l$ in $R$, in which $P(q_l) = \frac{|\{r_k \in R : q_{r_k} \equiv q_l \}|}{|R|}$. 

$P(\tilde{c_i})$ is the probability of a random selected query that contains $\tilde{c_i}$ as one of its attributes. It is computed from (4):
\begin{equation}
P(\tilde{c_i}) = \sum_{r_k \in R} \left( P(\tilde{c_i}|q_{r_k})\cdot P(q_{r_k}) \right)
\end{equation}
By considering all the involved context, the probability of $q_l$ (denoted by $P(q_l|\tilde{C}, R)$ will be refined as (5):
\begin{equation}
P(q_l|\tilde{C}, R)=
\sum_{\tilde{c_i} \in \tilde{C}}
\left( 
\frac{P(\tilde{c_i}|q_l) \cdot P(q_l)}{P(\tilde{c_i}) }
\cdot
\frac{1}{ |  \tilde{C} | } 
\right)
\end{equation}
The calculation from (5) is based on considering the importance of all involved contexts equally. However, the importance of each context must be distinguished by different users. Hence, we apply the \textit{weight of context} \cite{DelirHaghighi2008} in our scheme. \newline

\noindent \textbf{Definition 5: \textit{Context Importance Rules---$G$}.} $G$ is a finite set of rules, where $G = \{g_m : 1 \leq m \leq N\}$. Each $g_m$ consists of a corresponding context $c^{g_m}$ and a corresponding query $q^{g_m}$, and the weight value denoted by $v_{g_m}$. 

$v_{g_m}$ is a user-defined value in the context importance rules ($G$) for clarifying the importance of a context type to a query. By default setting, each context type has equal importance (set to 0) to all the queries. For example, a user may consider the location context to be more important to a query for searching the train arrival time. Hence, the user can increase the importance of the location context (e.g., set it to a number greater than zero) to the query to improve the prediction accuracy. Such a setting can also be applied globally. For example, user may prefer the location context should always be the primary consideration. Hence, whenever the prediction is performed, the location context will always be allocated a higher importance value than the other contexts.   

By applying the \textit{weight of context}, the final formula has been refined in (6).

\begin{equation}
P(q_l|\tilde{C}, R)=
\sum_{\tilde{c_i} \in \tilde{C}}
\left( 
\frac{P(\tilde{c_i}|q_l) \cdot P(q_l)}{P(\tilde{c_i}) }
\cdot
\frac{1+v_{g_m}^{c_i, q_l}}{ |  \tilde{C} | + \sum v_{g_m}} 
\right)
\end{equation}
where $\sum v_{g_m}$ is the sum of a set of $v_{g_m}$ in which $v_{g_m} \neq 0$. $v_{g_m}^{c_i, q_l}$ denotes one of the defined rule, where $v_{g_m}^{c_i, q_l} \leftarrow g_m \in G, c^{g_m} \equiv \tilde{c_i} \wedge q^{g_m} \equiv q_l$.

In order to let a user have enough control on the autonomous decision-making based on the prediction model, the user can manually define Context Filtering rules. A Context Filtering rule consists of a query type, and a list of contexts that should be ignored in the calculation. For example: A user is searching for the recommended food in the current area. For this search query, weather context and temperature context can be important if the \textit{food seller type} is an outdoor bazaar, or it does not have enough indoor seats when customers are required to queue outside. On the other hand, a similar search method may not be influenced by weather and temperature if the query specifies the search criteria as ``\texttt{restaurant}" + ``\texttt{indoor}".

If user defined rules exist, in the prediction algorithm, the current contexts $\tilde{C}$ for a query $q_l$  will be redefined to reflect whether a context should influence the $q_l$ or not. For example, current contexts $\tilde{C}$ consists of $\tilde{c_1}$, $\tilde{c_2}$, $\tilde{c_3}$, and $\tilde{c_4}$. If user has defined that $\tilde{c_4}$ has no influence to \textit{query type} $q_y$, when the prediction algorithm computes the $P(q_y|\tilde{C}, R)$ (see (6)), $\tilde{C}$ will be redefined as $\left\{\tilde{c_1}, \tilde{c_2}, \tilde{c_3} \right\}$ excluding $\tilde{c_4}$. 

A prediction scheme that relies on the user's historical record usually has a limitation in which the  accuracy of the prediction can be low when there is not enough records. One solution is to apply \textit{social context}. \textit{Social context} represents the factors that can potentially influence a user's decision. For example, a friend $f$ of a mobile user $u$, might have similar interest to $u$, and $f$ might have been to the same place as where $u$ is currently arriving. Since $f$ and $u$ are similar, they may prefer to interact with the same type of services at that location. 

%% file: Trust.tex
\section{Trustworthy Service Discovery for MSNP}
This section presents a scheme to improve the speed of trustworthy service discovery in service-oriented MSNP by reducing transaction overhead and not relying on message forwarding in order to avoid the issues caused by unstable connectivity and resource constraint.

\subsection{A Lightweight Trustworthy Service Discovery for MSNP}
The fundamental strategy to reduce the transaction overhead in MSNP is to utilise the selective trust reputation rating recommender scheme similar to existing works. However, we need to address two additional issues:
\begin{enumerate}[(1)]
\item How can MSNP participants share their reputation rating data?
\item How can a requester limit the number of its recommenders in the friend-based reputation model and in a public-based reputation model?
\end{enumerate}
The later sections involve a number of elements. Hence, we define the meaning of the elements first.
\newline\newline
\noindent \textbf{Definition 6: \textit{Service Provider}---$SP$.} An $SP$ is an MSNP participant that provides WS. It is defined as a tuple $(ID, services)$ where:
	\begin{itemize}
	\item[---] $ID$ denotes the identity of the $SP$
	\item[---] $services = \{service_i : 1 \leq i \leq \mathds{N} \}$ represents a set of WS provided by the $SP$. Each $service$ has a name denoted by \textit{SName} and a semantic service type denoted by \textit{SType}
	\end{itemize}

\noindent \textbf{Definition 7: \textit{Previous Interacted Service Consumers List}---$PSC$ \textit{list}.} $PSC=\{(cid_j, IR_j) : 1 \leq j \leq \mathds{N} \}$. 
An $SP$ can optionally provide its $PSC$ list to let the others know who have been using its services. A $PSC$ is defined as a tuple $(cid, IR)$ where:
	\begin{itemize}
	\item[---] $cid$ denotes a service consumers' identity
	\item[---] $IR$ denotes interaction records between the service provider and service consumer, e.g., $IR_j$ denotes a list of interaction records between the $SP$ and the service consumer $cid_j$
	\end{itemize}

\noindent \textbf{Definition 8: \textit{Service Provider Ratings}---$SPR$.}$SPR = \{(ID_k, Rates_k) : 1 \leq k \leq \mathds{N}\}$ where:
	\begin{itemize}
	\item[---] $ID_k$ denotes the identification of $SP_k$
	\item[---] $Rates_k = \{(service_l^k, rate_l^k) : 1 \leq l \leq \mathds{N}\}$ is a list of rating values of $SP_k$'s services
	\item[---] $service_l^k$ denotes one of the $SP_k$'s services
	\item[---] $rate_l^k$ denotes the rating value of $service_l^k$ 
	\end{itemize}

\noindent \textbf{Definition 9: \textit{Recommended References}---$RR$.} $RR = \{(SType_m, ID_m) : 1 \leq m \leq \mathds{N}\}$ where:
	\begin{itemize}
	\item[---] $SType$ denotes a semantic service type
	\item[---] $ID_m = \{id_o^m : 1 \leq o \leq \mathds{N}\}$ denotes a list of MSNP participants' IDs that are recommended as the rating reference for $SType_m$ services
	\end{itemize}

\noindent \textbf{Definition 10: \textit{Reputation Rating Data}---$RD$.} Each MSNP participant has a $RD$ file in its device local storage as well as its cloud storage synchronously. An $RD$ file contains two sets of data---$SPR$ and $RR$.  
 
\begin{lstlisting}[basicstyle=\footnotesize\ttfamily,label=some-code,caption=Simplified $RD$ example]
<key>Service Provider Rating</key>
<value>
 <key>SPID</key>
 <value>
  <key>URI</key>
  <value>
   <key>type</key><value>semantic type value</value>
   <key>Rate</key><value>rating value</value>
   <key>transaction records</key>
   <value><!--URI, service type, time etc.--></value>
  </value>
</value>
<!-- Other interaction records ... -->	
</value>
<key>Recommended References</key>
<value>
 <key>Semantic Service Type</key>
 <value>
  <key>ID</key><value><!--URL of RD--></value>
  <!-- other IDs ... -->
 </value>
 <!-- Other types ... -->	
</value>
\end{lstlisting}

Listing 1 illustrates a simplified $RD$ in hash map format. An $RD$ file can be obtained from either friends or other proximal MSNP participants. The prerequisite condition is how the requester \textit{agent} retrieves the $RD$ from the other \textit{agents} (either from friends or public proximal participants). In a generic Mobile Ad Hoc Network environment, it is commonly assumed that the requester \textit{agent} will collect the $RD$ by broadcasting or multicasting its request message to the other participants' \textit{agents}. This is not always applicable in MSNP. Fundamentally, MSNP operates in a dynamic public MP2P environment in which participants may not always be available. For example, when the requester \textit{agent} intends to request a list of friends' \textit{agents} for the $RD$, there may only be a few of them online. Another example, when the requester \textit{agent} intends to request a list of proximal MSNP participants' \textit{agents} for the reputation rating data, many may not even respond to such a request because they may have disabled such an operation to save their battery power.  

To resolve the basic data retrieval problem in MSNP, each MSNP participant can utilise one or multiple backend public accessible cloud storage services to provide its $RD$ to the others. The URL of the $RD$ can be simply described in SDM. Hence, while the requester \textit{agent} retrieves Service Description Metadata (SDM; e.g., WSDL) in the first phase of service discovery process, it can already identify where to retrieve the reputation rating data provided by the other proximal participants. As for the friends' $RD$, since the requester has close connection with them, the requester would have already replicated their SDM files. Therefore, the requester \textit{agent} always knows where to retrieve the $RD$ of the requester's friends.

One aspect in MP2P trust that was not addressed in most existing works is how service providers actively participate in the trustworthy service discovery processes. In real world services, providers always attempt to encourage consumers to use their services by using various schemes such as showing customers' rating and reviews of their products and services. Although in an MP2P trust system, service providers should not hold the rating of their own services \cite{Singh2003}, they can still provide a list of previous interacting service consumers. 

When a requester intends to retrieve a service provider's reputation rating, the service provider can provide a $PSC$ list. The requester can use the $cid$ of $PSC$ list to collect \textit{RD} instead of collecting all the \textit{RD} of friends or proximal strangers. This approach can reduce unnecessary data transmission. Moreover, MSNP \textit{agents} can identify that a service provider who does not provide the $PSC$ list can potentially be a malicious node unless the service provider is new to the MSNP. If an MSNP participant is new, it may not have any interaction record with any other participants either as a service consumer or as a service provider. Hence, if an MSNP participant is not new, and it does not provide the $PSC$ list, then this MSNP participant's service can be identified as potentially malicious. This notion is based on the reputation system of general online trading/shopping services such as eBay\footnote{See \url{http://www.ebay.com/}} or Yahoo Auction Japan\footnote{See \url{http://auctions.yahoo.co.jp/}}. In the case of only one matched service provider found in the network, and it is a new MSNP participant, the \textit{agent} should notify its user and let the user decide whether to invoke the service or not. 

Considering the situation when a dishonoured service provider may provide an incomplete $PSC$ list, which only describes a list of good records, the requester \textit{agent} should not refer to the service provider's $PSC$ list to identify the service provider's trustworthiness in the following cases: 

\begin{itemize}
\item In the case of recommendation from friends: If none of the $cid$ found in $PSC$ belongs to the requester's trusted friends, the $PSC$ should not be used.
\item In the case of recommendation from public: If none of the $cid$ found in the $PSC$ belongs to highly creditable strangers, the $PSC$ should not be used.
\end{itemize}

The following sections describe the proposed scheme for trustworthy service discovery in service-oriented MSNP.


\subsection[Selecting Recommenders Based on Friends and FOAF]{Selecting Recommenders Based on Friends and FOAF}

Due to privacy issues, the information about a person's trust rating value to his/her friends may not be accessible to other friends. For example, in Facebook, a user can hide all their posts from a friend and the friend will not know. Although the trust rating value is not accessible, the person can still provide a list of friends as $RR$ for a particular service type. The friends' Identifications (IDs) assigned in $RR$ denote that the owner of $RR$ trusts this list of participants' judgement for a particular service type based on their past experience.

$RR$ is generated and updated when an MSNP \textit{agent} performs service by referring to the \textit{RD} of its user. $RR$ only contains the IDs of trusted friends for a particular service type. If a friend in this list has given a high rating to a bad service provider, the friend's ID will be removed from the list. As a simple example, the MSNP application lets user manually block a service provider ID. When a service provider ID is blocked, the friends who gave a good rate to the service provider will be removed from the corresponding Recommended References. On the other hand, when the list is empty and the recommendation was from randomly picked friend, if a friend's recommended service provider gives satisfactory recommendation to the requester, the friend's ID would be added to the list.\newline

\noindent There are two approaches to assign friends to $RR$: 
\begin{enumerate}[(1)]
\item \textit{Based on experience}. Since an $RD$ provides a list of ratings, an \textit{agent} is capable of identifying which friend of its user has the highest service interaction experience with a specific service type. 
\item \textit{Based on similarity}. A user can assign their friends to $RR$ based on how similar their past rating to a particular service type. i.e., using Pearson Product-moment Correlation Coefficient. For example, \textit{user A}'s past rating is very similar to \textit{user B}. \textit{user C} intends to refer \textit{user A}'s rating to service provider---$H$ who provides $K$ type service. Unfortunately, \textit{user A} does not have experience with $H$. However, since \textit{user B}'s rating is similar to \textit{user A}, \textit{user A} has already assigned \textit{user B} as $RR$ of $K$ type service. Hence, \textit{user C} will refer to \textit{user B}'s rating to identify the trustworthiness of $H$. 
\end{enumerate}

The rating seminaries between two users---$A$ and $B$---can be computed by using Pearson Product-moment Correlation Coefficient below:


\begin{align}
sim(A, B) =  
\frac{ 
	\sum_{s\in S} (  rate_{A \rightarrow s} - \overline{rate}_r  )
	(  rate_{B \rightarrow s} - \overline{rate}_{B}  )
}
{ 
	\sqrt{  \sum_{s\in S} ( rate_{A \rightarrow s} - \overline{rate}_A  )^2   \sum_{s\in S} ( rate_{B \rightarrow s} - \overline{rate}_{B}  )^2 }
} 
\end{align}
\\where $S$ is the set of all the services rated by both $A$ and $B$. $rate_{A \rightarrow s}$  is $A$'s rating to service $s \in S$. $\overline{rate}_A$ is the average rating rated by $A$ to all services. $rate_{B \rightarrow s}$ is $B$'s rating to $s \in S$, and $\overline{rate}_{B}$ is the average  rating rated by $B$ to all $services$.

Note that both approaches require a fair number of friends' \textit{RD} replicated previously. For example, a user can replicate their friends' \textit{RD} at home, then their \textit{agents} can apply the approaches to identify $RR$ before the user using MSNP application outside. 

The following algorithm outlines the steps for a requester to identify the trust score of a service/content provider's service  $s \in S$. \newline

\noindent \textbf{Algorithm 1:}

\noindent \textbf{Step 1.} \textit{Identify a list of friends who have experience with service}---$s$. 
\begin{itemize}
\item[1.1.] Requester retrieves $PSC$ of the provider of $s$  ($PSC_{s}$). We expect that the requester has a list of friends' IDs (denoted by $FID$, where $FID = \{ fid_j : 1\leq j \leq N \}$) stored in the local memory of the mobile device. 
\item[1.2.] By searching the intersection between all the $cid$ in $PSC_s$ and $FID$, requester can find a list of friends who have service invocation experience with $s$---$MFID$, where $MFID=FID\cap CID$. If $|MFID|=0$ then the process goes to Step 3. Otherwise, continue with Step 2.
\end{itemize}

\noindent \textbf{Step 2.} \textit{Identify matched recommended references}. 
\begin{itemize}
\item[2.1.] As described previously, each MSNP participant has a \textit{RD}. Let $MRR = \{ rr \in RR : SType_{rr} \equiv SType_{s} \}$, where $SType_{s}$ is the semantic service type of $s$ that the requester intends to invoke. $RR$ is a list of friends' IDs that are recommended for identifying the reputation of a type of $SType_{s}$. 
\item[2.2.] Let $RrID = MFID \cap MRR$. From $RrID$, the requester \textit{agent} can identify the recommended friend(s) for $SType_{s}$ that also have experience with $s$, and refer the friend's rating to $s$. If $|RrID| = 0$, the process goes to Step 3. 
\end{itemize}

\noindent \textbf{Step 3:} \textit{Referring recommendation from recommended friend's FOAF}. When the requester's direct friends do not have experience with $s$, the requester will refer to the reputation rating from FOAF. 
\begin{itemize}
\item[3.1.] Identify a friend with the highest experience  as a recommender and then based on the recommender's $RD$ to find the friend of the recommender who has the highest experience with $SType_{s}$ and who also has rated $s$. 
\item[3.2.] Once the FOAF is found, the requester will refer to the FOAF's rating of $s$. However, if none of the FOAF has experience with $s$ then the process will proceed to the scheme described in Section 6.3---Selecting recommenders based on public. 
\end{itemize}


\subsection{Selecting Recommenders based on the Public}

In this section, we describe the scheme to identify a service provider $SP$'s reputation score based on the public proximal MSNP participants' ratings. 
\newline\newline
\noindent \textbf{Definition 11: \textit{Credibility}---$Cr$.} An MSNP participant's $Cr$, which is rated by the other peers, represents its reputation as a recommender for a type of service. The more MSNP participants' IDs shows up in the $RR$ of every peer's \textit{RD}s, the higher the MSNP participant's credibility is for being a recommender of the corresponding service type. \newline

Algorithm 2 describes the scheme for selecting trustworthy recommenders from public proximal MSNP participants. \newline

\noindent \textbf{Algorithm 2:}

\noindent \textbf{Step 1:} \textit{Generating a candidate recommender list}. 
While the requester performs the service discovery process to find service providers who can provide the service of interest, the requester is also retrieving the \textit{RD} of each proximal MSNP \textit{agent}. This step consists of the following two tasks: 
\begin{itemize}
\item[1.1.] Let \textit{PRRD} be the set of \textit{RD}s retrieved from all proximal \textit{agents}. $PRRD=\{prrd_i : 1 \leq i \leq N\}$  where $prrd_i$ denotes the $RD$ of each \textit{agent} $p_i$. For each $prrd \in PRRD$, the requester \textit{agent} can identify that whether a $p_i$ has interaction experience with service provider $s$ or not.
\item[1.2.] Let $MPR$ denotes the matched $PRRD$ in which $MPR=\{prrd_j \in PRRD | ID \in SPR_j \equiv ID_s\}$. $ID_s$ denotes the $ID$ of service provider $s$. If $ID_s$ is found in one of $prrd_j$'s $SPR$ but not in the $PSC$ list of the provider of $s$, then either the $prrd_j$ is dishonoured or the provider of $s$ is dishonoured. 
\end{itemize}
Since the aim of this scheme is to identify the trust of $s$'s provider, the final result will show its reputation score. However, dishonoured rating from the other participants will affect the accuracy of the scheme. Hence, the requester \textit{agent} has to identify a recommender's trustworthiness before referring its reputation rating. Step 2 describes the process to identify a recommender's trustworthiness based on credibility.

\noindent \textbf{Step 2:} \textit{Identify the credibility of a candidate recommender}. 
A proximal MSNP participant's credibility is computed based on the other proximal MSNP participant's rating. Suppose we want to compute a proximal MSNP participant $p_i$'s credibility, we will use $PRRD$ excluding the \textit{RD} of $p_i$.  We use $CRRD$ to represents such a set of data, where $CRRD=\{ crrd_m : 1 \leq m \leq N \}$. Step 2 consists of the following two tasks:
\begin{itemize}
\item[2.1.] Let $Cr_{p}$  be the credibility of $p$, and $Cr_{p}$ is computed by the formula below:  
\begin{equation}
Cr_{p}=|\{crrd \in CRRD | ID_{rr_o^{crrd}} \equiv ID_{p} \}|
\end{equation}
where  $ID_{rr_o^{crrd_m}}$ denotes an MSNP participant's ID in the $RR$ of $crrd_m$, and $ID_{p}$ denotes $p$'s ID in MSNP. 
\item[2.2.] Once the credibility of each $PRRD$'s owner $p_i$ is computed, the process goes to the next step.
\end{itemize}

\noindent \textbf{Step 3:} \textit{Identify the experience of a candidate recommender}.
People trust a person who has more experience about a specific subject. In existing works such as TEMPR \cite{Waluyo2012}, the experience of $p$ is directly related to the number of successful interactions completed between $p$ and the service provider. Here, we consider the experience based on the type of service instead of a particular service provider's service. Because in the real world, a person may not use a service the second time when he/she had a bad experience with the service the first time. However, the person may have a lot of of experience using the same type of service provided by many different providers. Hence, the person's opinion is still valuable. For example, the review of a senior computer machine reviewer, who has over 100 reviews of notebook computers from different brands, is often being considered as more trustable than a junior reviewer who has only reviewed less than 10 notebook computers.  Based on this assumption, the experience of $p$ in our model is based on $p$'s experience to a particular service type. This step involves the task below:
\begin{itemize}
\item[3.1.] Let  $STypeEx_{p_i \rightarrow s}$ be $p_i$'s experience to $SType_s$. The experience value of $p_i$ to $SType_s$ is computed by:
\begin{align}
STypeEx_{p_i\rightarrow s} =  
|\{  ir_l^{RD_{p_i}} \in IR^{RD_{p_i}} : SType_{ir_l}^{RD_{p_i}} \equiv SType_{s}  \}|
\end{align}
\\where 
	\begin{itemize}
	\item[---] $IR^{RD_{p_i}}$ is the interaction records of $p_i$, in which \\$IR^{RD_{p_i}} = \{ ir_l^{RD_{p_i}} : 1 \leq l \leq N  \}$. 
	\item[---] $SType_{ir_l}^{RD_{p_i}}$ denotes the service type of the invoked service recorded in $ir_l^{RD_{p_i}}$.
	\end{itemize}
\end{itemize}

\noindent \textbf{Step 4:} \textit{Compute the trust score of a candidate recommender}.
The trust score of an MSNP participant is the average of its normalised credibility value and its normalised experience value. The normalised value is computed based on the overall comparison from all the other participants in $P$. This step involves the following two tasks:
\begin{itemize}
\item[4.1.] For a particular MSNP participant---$\varphi \in P$  as a recommender of a service type ($Tr$), the trust score $Tr_{\varphi}$ of $\varphi$ is computed by the formula: 
\begin{equation}
Tr_{\varphi} = avg\left (
\frac{ 
	Cr_{\varphi}
}
{ 
	\sum_{p_i\in P} Cr_{p_i}
}
+
\frac{
	STypeEx_{\varphi \rightarrow s}
}
{
	\sum_{p_i\in P} STypeEx_{p_i \rightarrow s}
}
\right )
\end{equation}
where 
	\begin{itemize}
	\item[---] $Cr_{\varphi}$ is $\varphi$'s credibility value. 
	\item[---] $\sum_{p_i\in P} Cr_{p_i}$ denotes the sum of  credibility values of all $p_i$. 
	\item[---] $STypeEx_{\varphi \rightarrow s}$ denotes the experience of $\varphi$ for $SType_s$. 
	\item[---] $\sum_{p_i\in P} STypeEx_{p_i \rightarrow s}$ denotes the sum of all $p_i$'s experience for $SType_s$.
	\end{itemize}
	
\item[4.2.] Based on the computation result, the requester can choose a number of MSNP participants that have the highest $Tr_{\varphi}$ value to be its recommender to compute the reputation score of $s$.
\end{itemize}

%% file: Evaluation.tex
\section{Evaluation}
For proof-of-concept, we have developed and evaluated a prototype consisting of the components composing the mechanism described in our schemes. This section presents the evaluation methods and results of our prototype. In order to show the detailed evaluation of each proposed scheme, we have tested each component individually.

\subsection{Prototype Implementation}
The prototype was developed using the objective-C programming language and was tested on Apple iPod Touch 4th generation and Apple iPhone4S. 

The basic mechanism which  lets MSNP \textit{agents} participant in the service-oriented MSNP is Mobile Web Service (MWS). Depending on the user preference, an \textit{agent} can either support the simple MWS client-side mechanism only to discover and invoke services or support both MWS client-side and MWS server-side mechanisms. The implementation of the MWS mechanism are described below:

The MWS provided by each MSNP \textit{agent} in the prototype is RESTful Web service. An advanced MSNP content consumer or a content provider is able to be an MWS host. In the prototype, an MWS host consists of two main components:
\begin{itemize}
\item \textit{\textbf{HTTP Web server}}\\We used CocoaHTTPServer\footnote{\url{https://github.com/robbiehanson/CocoaHTTPServer}} API to enable the HTTP server mechanism. The advantages of using CocoaHTTPServer are: (1) it supports asynchronous socket communication, which can improve the speed of data transaction; (2) it supports Bonjour service publication. Web services provided by CocoaHTTPServer are discoverable by the Bonjour service discovering mechanisms. In the prototype, we use Bonjour as the main mechanism for MP2P service discovery. The HTTP Web server in prototype will respond with a SAWSDL document when the request message does not specify a particular path/operation name.
\item \textit{\textbf{Semantic Web service protocol}}\\SAWSDL and OWL documents play an important role in MSNP. In order to enable autonomous service discovery and filtering, an MWS host is required to be able to process XML-formatted SAWSDL and OWL. We used Google Data API\footnote{\url{https://developers.google.com/gdata/}} to process XML-formatted data. Google Data API provides a fully functioned XML parsing mechanism. The SAWSDL and OWL data used in the prototype were written manually because there is no tool available to generate SAWSDL automatically from the source code written in Objective-C. 
\end{itemize}

The basic functions of an MSNP \textit{agent} are: to discover other MSNP \textit{agents} in its current network; and to invoke Web services provided by the other \textit{agents}. In order to support the two functions, we have implemented two components:
	\begin{itemize}
	\item \textit{\textbf{Web service invocation component}}\\It supports two asynchronous HTTP method invocation mechanisms (\texttt{GET} and \texttt{POST}), which is compatible with RESTful Web services.
	\item \textit{\textbf{MP2P service discovery component}}\\The prototype used Bonjour technology to support MP2P service discovery. Each MSNP \textit{agent} has a Service Pool component to monitor the current network. The Service Pool component utilises  \texttt{<NSNetServiceBrowserDelegate>} to monitor the published MWS (by MSNP \textit{agent}) in the Bonjour network. It manages a list of pushed MWS names. Depending on the discovery approach, it may automatically retrieve the SDM of each newly joined MSNP using the Web service invocation component. 
	\end{itemize}

Evaluating the performance of service discovery may involve hundreds of MSNP \textit{agents}.  We  did not have a large number of mobile devices to realise such an  environment. However we have deployed hundreds of MSNP agents in a Macbook Aluminium 2008 version with Intel Core 2 Due 2.4 GHz CPU and 4GM RAM to simulate the environment. The wireless network for evaluation is on IEEE 802.11n 2.4GHz Wi-Fi environment controlled by an Apple Airport router which is Internet connection-enabled.

The following sections provide details on how each component was evaluated in order to present the proof-of-concept of our proposed schemes.


\subsection{Proactive Service Discovery Performance}
In Section 4.3, the user preference associated push-based service discovery \\($PrefPush$) approach was described. The $PrefPush$-based service discovery approach utilises the context-aware user preference prediction scheme to let the requester $agent$ provides its user preferred semantic service type to other service/content provider $agents$ in the network. The approach can reduce the required metadata processing on the requester-side, hence, reducing the service discovery timespan of the requester $agent$. 

In order to show that the proposed $PrefPush$-based service discovery approach can provide a better performance than the other two basic approaches---$Pull$ and $Push$. We have performed an experiment in a simulation environment to compare the performance (timespan of  service discovery process) and the costs (CPU usage and RAM usage) of the three approaches.

First, we re-iterate and describe the implementation details of the three approaches below:

\begin{itemize}
\item $Pull$, as described in Section 4.1, enables an MSNP requester \textit{agent} who intends to search for a $cType$ service in an MSNP environment to use active invocation to retrieve the other service provider \textit{agents}' service description metadata (SDM) in order to identify which \textit{agent}(s) can provide the $cType$ service. In our setting, we assume each service provider \textit{agent} has its own OWL file to describe its semantic type. Hence, the requester \textit{agent} has to retrieve both SAWSDL and OWL from each service provider \textit{agent}. 
\item $Push$, as described in Section 4.2, allows the MSNP requester \textit{agent} to utilise MWS to passively receive and process SDMs advertised by the other service provider \textit{agents}. 
\item $PrefPush$, as described in Section 4.3, is fundamentally similar to the $Push$ approach. However, the requester \textit{agent} is also able to provide user preferred service type to the other service provider \textit{agents}. In this approach, the semantic type parsing process is performed by the service provider \textit{agents}. 
\end{itemize}

In our experiment, we did not include the $Hybrid$ approach (described in Section 4.4) in the comparison because the main purpose of the $Hybrid$ approach, which utilises both $Pull$ and $PrefPush$ concurrently, is to guarantee that the requester \textit{agent} is still capable of performing service discovery in an MSNP environment when the other \textit{agents} do not support the $PrefPush$-based approach. In other words, it provides a fall back mechanism.

\subsubsection{Settings}

The experiment was performed mainly on an iPhone4S with IEEE802.11n 2.4 GHz Wi-Fi environment. We simulated the other proximal MSNP participants by deploying a number of MSNP agent hosts on a Macbook. 

The MSNP \textit{agent}, which has been installed in the iPhone4S, represents the service requester who is searching for MSNP \textit{agents} who can provide the service that match a semantic service type---$cType$. The proximal MSNP participants' MSNP \textit{agents} are of two types: $normal$ and $matched$. The $normal$ MSNP \textit{agents} do not provide $cType$ services and the $matched$ MSNP \textit{agents} can provide $cType$ services. Every MSNP \textit{agent} (including the requester and the others) provides two service description related documents---SAWSDL and OWL. The size of SAWSDL is 6KB and the size of OWL is 12KB. 

The experiment consists of two tests: performance and resource costs. 
\begin{itemize}
\item For the performance, we aimed to compare the timespans to successfully discover matched service providers from all the deployed MSNP \textit{agents} in the network. We deployed a different number of MSNP \textit{agents} on the Macbook to evaluate the three different approaches: $Pull$, $Push$ and $PrefPush$. 
\item For the resource cost testing, we aimed to compare the CPU and RAM usages between the three approaches. It was done by recording the CPU and RAM usages while performing the service discovery processes.
\end{itemize}
Note that in the description of Section 4.1, we have mentioned SDM caching. SDM caching can reduce transaction overheads for all approaches equally if implemented. Since the aim of our tests is to compare the three approaches solely, we did not include the caching mechanism. 

\subsubsection{Performance Comparison}

This section presents the experimental result of the service discovery timespan comparison among the three approaches, as shown in Figure 6.

\begin{figure}[!h]
  \centering
    \includegraphics[width=0.65\textwidth]{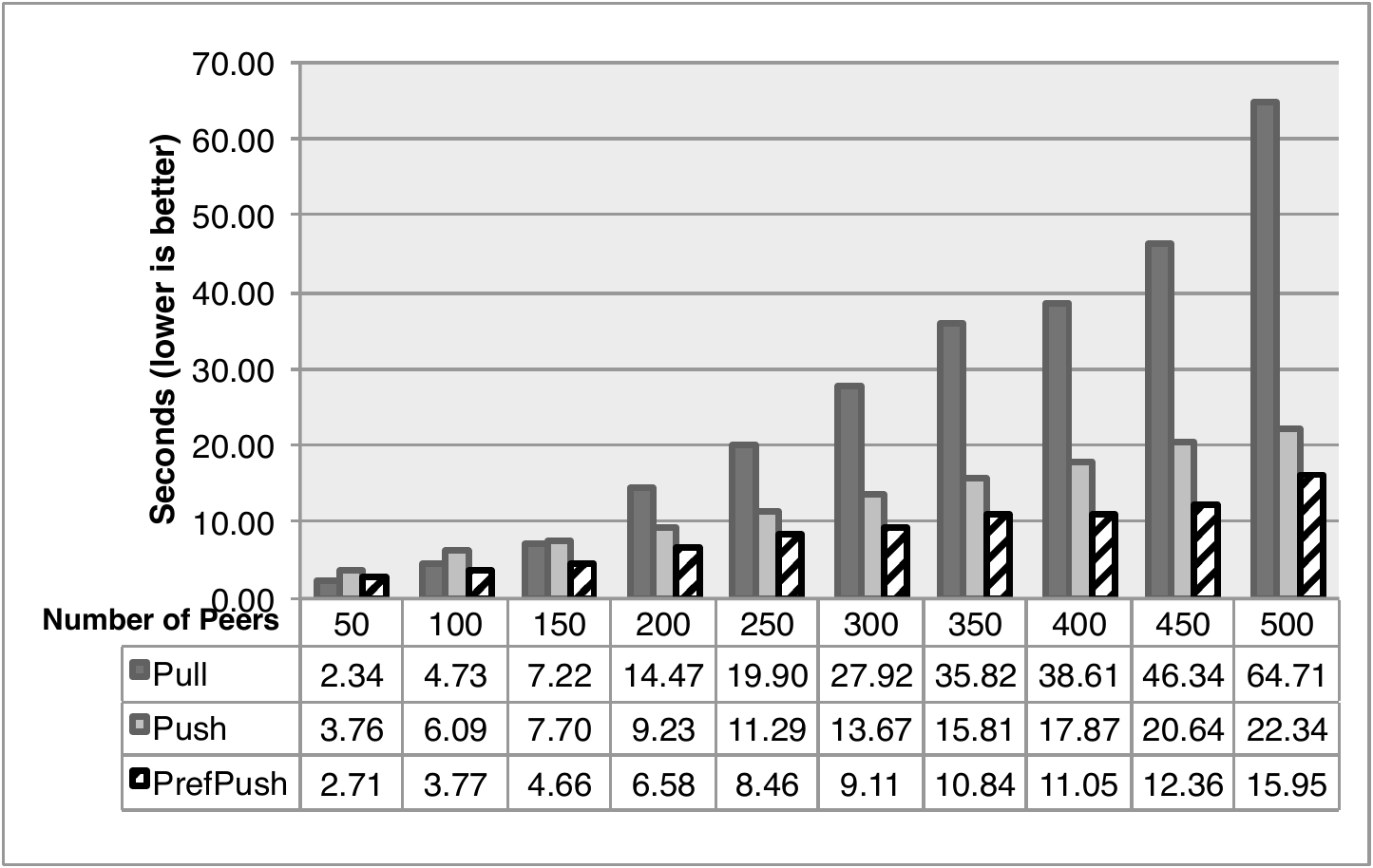}
    \caption{Timespan Comparison}
\end{figure}

In the figure, x-axis represents the number of service provider \textit{agents} deployed in the network. Each deployed group has 4/5 $normal$ \textit{agents} and 1/5 $matched$ \textit{agents}. For example, when 500 service provider \textit{agents} were deployed, while 400 out of 500 were normal agent, 100 out of 500 \textit{agents} were $matched$ \textit{agents} who can provide $cType$ service. The y-axis represents how long it took the requester \textit{agent} to discover the matched service providers. 

The result shows that when there were only 50 service provider \textit{agents} in the network, $Pull$ provided the best performance. However, when the number of service provider \textit{agents} increased, the performance of $Pull$ worsened  because of the increased amount of SDM retrieval and semantic data processing. $Push$ utilised MWS to receive SDM from the other service provider \textit{agents}. Although the requester \textit{agent} in the $Push$ approach also had to process SDM, the overall timespan was much lesser than $Pull$ when the environment consisted of a large number of service provider \textit{agents}. Among the three approaches, our proposed $PrefPush$ approach outperformed the other two when the environment consisted of 100 or more service provider \textit{agents}. 

\subsubsection{Resource Usage Comparison}

This section presents the comparison of the CPU and RAM usages of the three approaches. In our setting, we deployed 1 $matched$ service provider \textit{agent} and 4 $normal$ service provider \textit{agents} every 1 second continuously for 100 seconds. We recorded both CPU usage and RAM usage within a period of 100 seconds.

\begin{figure}[!h]
  \centering
    \includegraphics[width=0.65\textwidth]{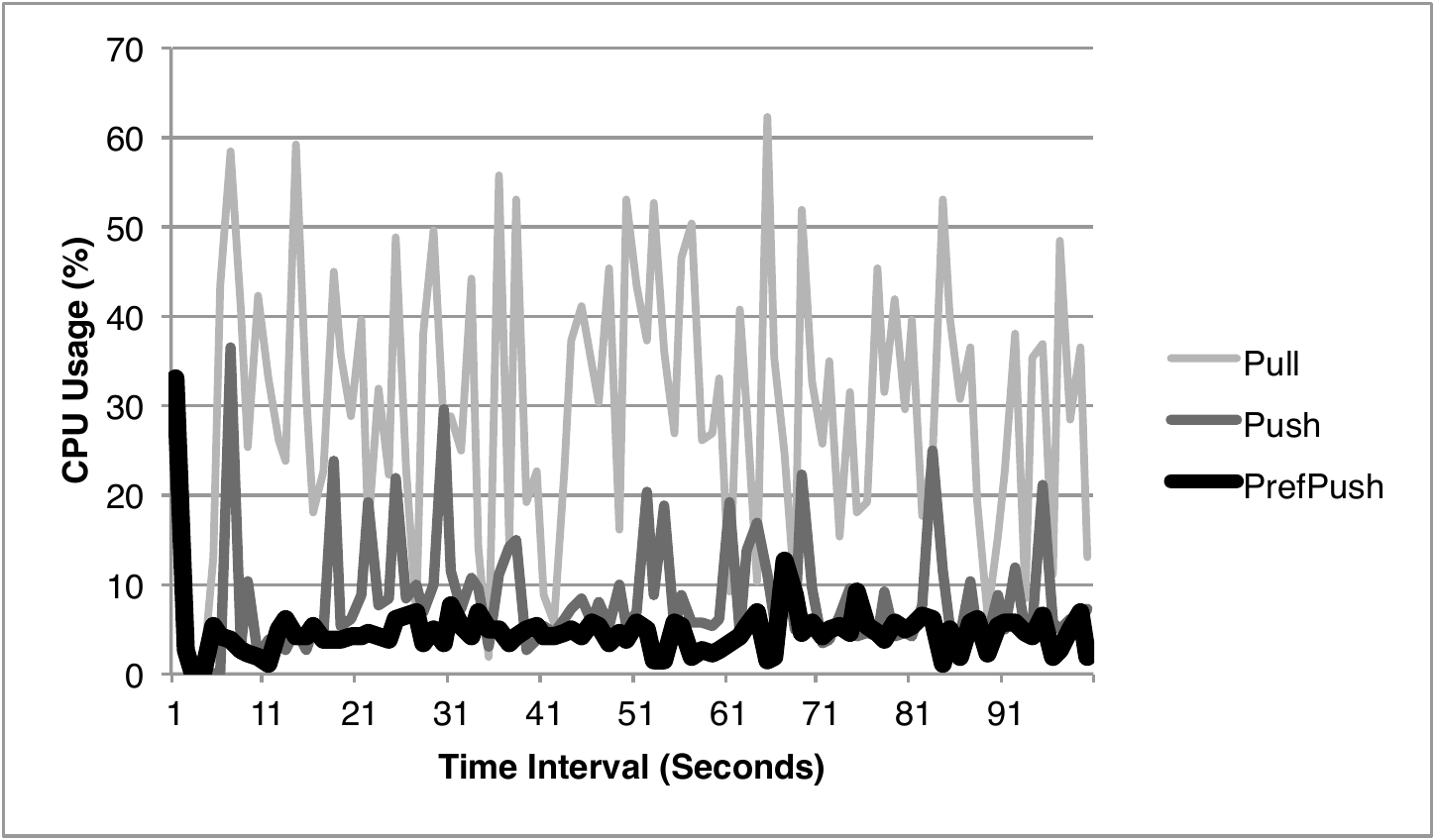}
    \caption{CPU Usage Comparison}
\end{figure}

Figure 7 illustrates the CPU usage record comparison among the three approaches. As the graph shows, $Push$ had the highest CPU usage while $PrefPush$ consumed the least CPU resource.

\begin{figure}[!h]
  \centering
    \includegraphics[width=0.65\textwidth]{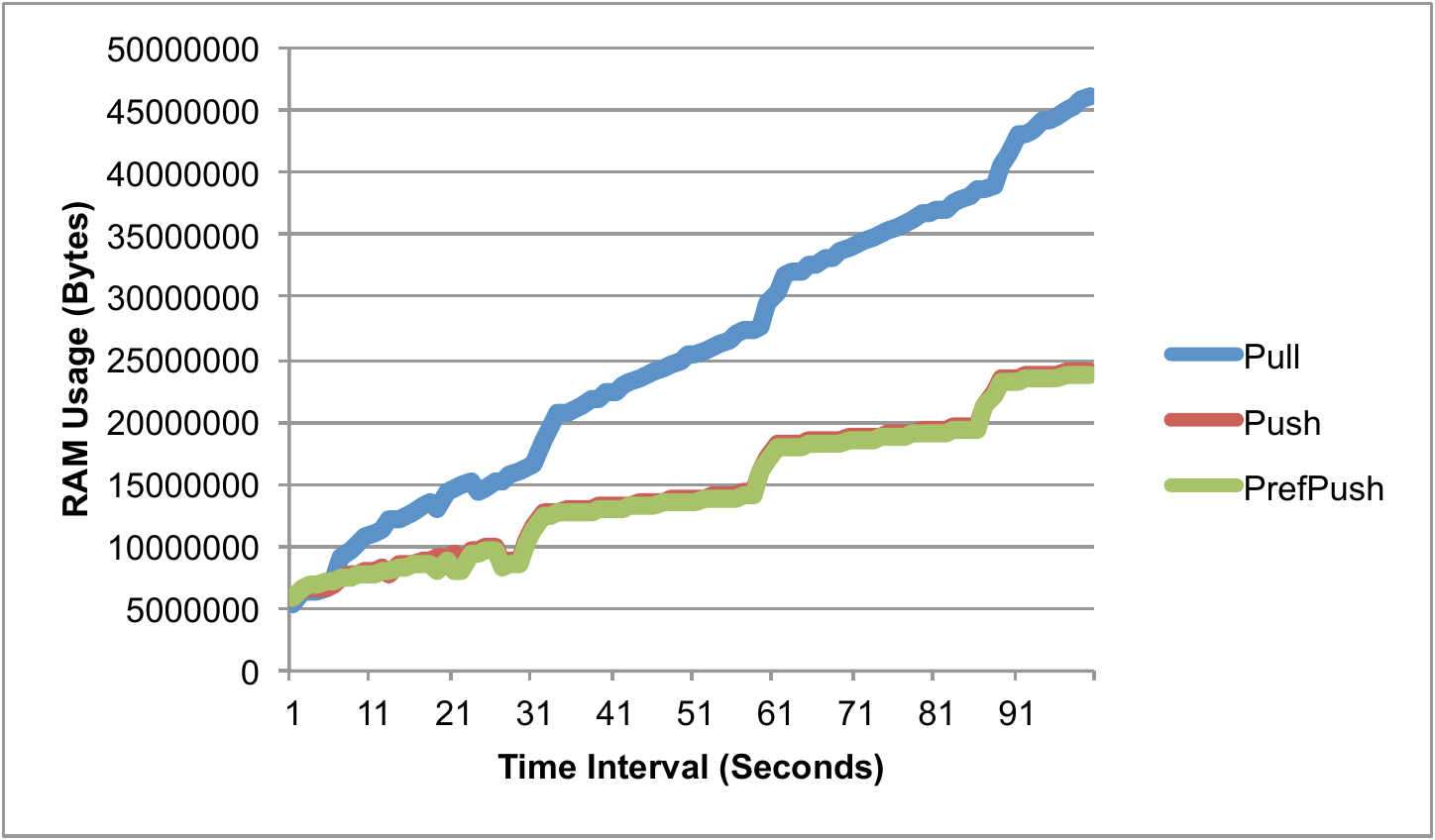}
    \caption{RAM Usage Comparison}
\end{figure}

Figure 8 illustrates the RAM usage comparison among the three approaches. As the result shows, the RAM usage of the three approaches increased continuously. $Pull$ consumed the highest RAM resource, while $Push$ and $PrefPush$ have very similar RAM resource consumption. 

In order to highlight the difference between $Push$ and $PrefPush$, we enlarged the graph to show the RAM usage between 40 to 60 seconds for $Push$ and $PrefPush$. This is shown in Figure 9.
\begin{figure}[!h]
  \centering
    \includegraphics[width=0.65\textwidth]{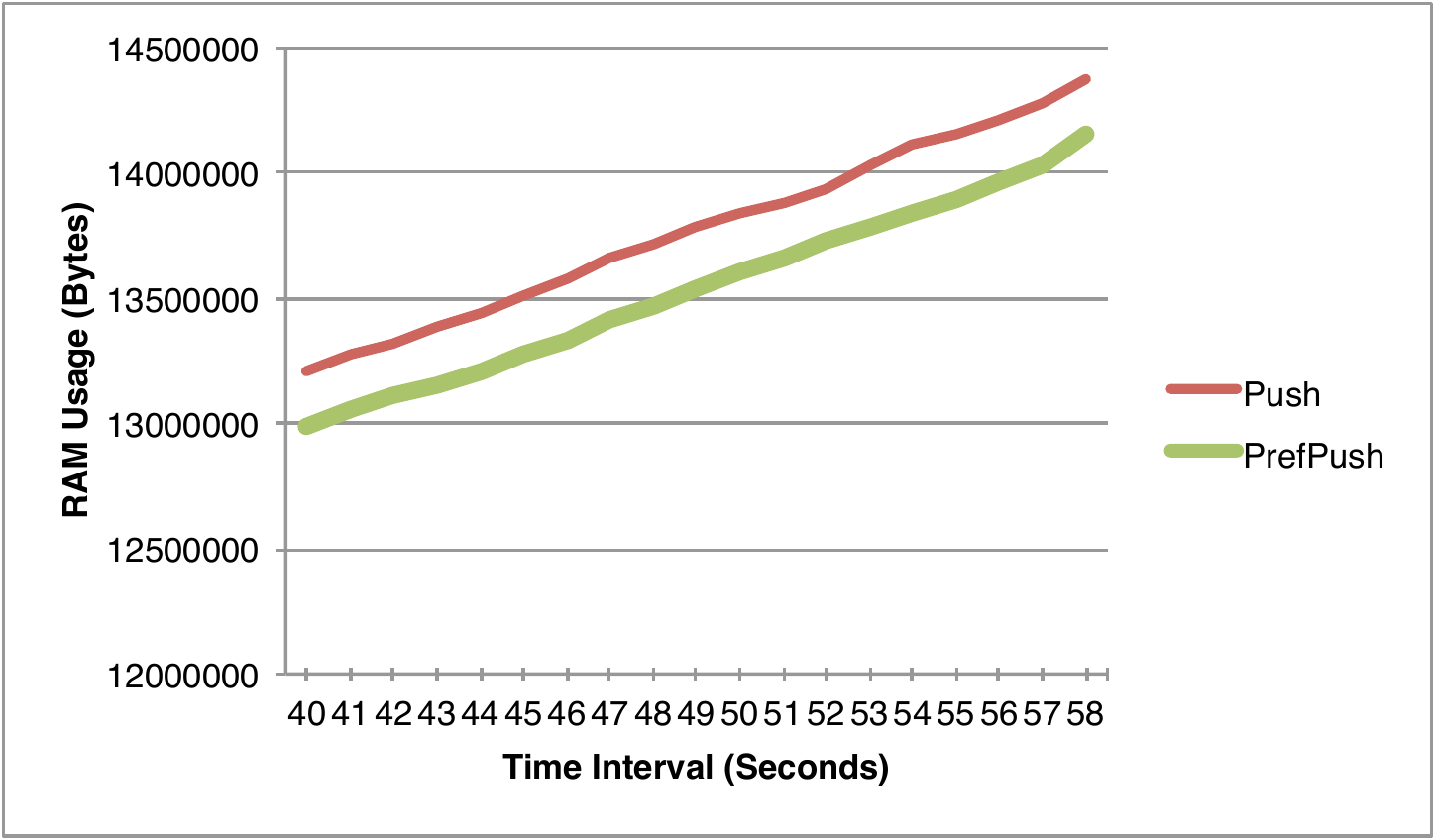}
    \caption{Partial RAM Usage Comparison of Push and PrefPush}
\end{figure}

In the figure, $Push$ shows a slightly higher RAM usage than $PrefPush$. 


\subsection{Context-Aware User Preference Prediction}
This section presents the evaluation results of the component that enables the context-aware user preference prediction scheme presented in Section 5.2, which is the main component for enabling the proactive service discovery in MSNP. The scheme uses current environmental context information to compare with the past context information associated with the service invocation records to predict what types of services may be of interest to the user in the current environment.

The test focused on evaluating the accuracy of the prediction scheme using two different datasets: the programme generated dataset and the epSICAR-dataset\footnote{\url{http://www.imada.sdu.dk/~gu/}}.

\subsubsection{Evaluating the Scheme on Programme \\Generated Dataset}
In order to test the accuracy of the scheme, we have created a user query record generator to simulate user query records and the associated context information. Table 1 illustrates the basic parameters used in the record generator. 
\begin{center}
\begin{table}[!h]
\centering
  \begin{tabular}{ | >{\small}p{1cm} | >{\small}p{1cm} | >{\small}p{1.4cm} | >{\small}p{1.4cm} | >{\small}p{1.4cm} | >{\small}p{1.4cm} | >{\small}p{1cm} |}
    \hline
     \textbf{Record} & \textbf{Query} & \textbf{CL} & \textbf{CT} & \textbf{CA} & \textbf{CW} & \textbf{CP} 
    \\ \hline \hline
   TypeA  & Q1 	& L1 		& T1	& A1-A5	& W1-W5	&P1-P5
    \\ \hline
    TypeB & Q2	&  L1-L5		& T2	& A2	& W1-W5	&P1-P5
    \\ \hline
   TypeC & Q3	&  L1-L5		& T1-T5	& A3 	& W3		&P1-P5
    \\ \hline
    TypeD &Q4	& L1-L5 		& T1-T5 & A1-A5 & W4		&P4
    \\ \hline
    TypeE & Q5	& L5		& T1-T5	& A1-A5& W1-W5		&P5
    \\ \hline
  \end{tabular}
  \caption{Parameters for Prediction Test}
\end{table}
\end{center}
We defined five types of records. Each record type describes a particular query type and five types of associated context information values denoted by CL, CT, CA, CW and CP. The record generator will randomly generate a given number of records (e.g., 100, 200, 300, etc.). Each record consists of one query type and five context values. For example, considering the setting in Table 1, the record generator will randomly select a record type from A to E. If the selected record type is A, then the query type will be Q1 and the associated context information will be CL=L1, CT=T1, CA = a random value from A1 to A5, CW = a random value from W1 to W5, CP = random value from P1 to P5. The two static values (L1 and T1) represent the contexts that will influence the user's decision to select Q1.

\begin{figure}[!h]
  \centering
    \includegraphics[width=0.65\textwidth]{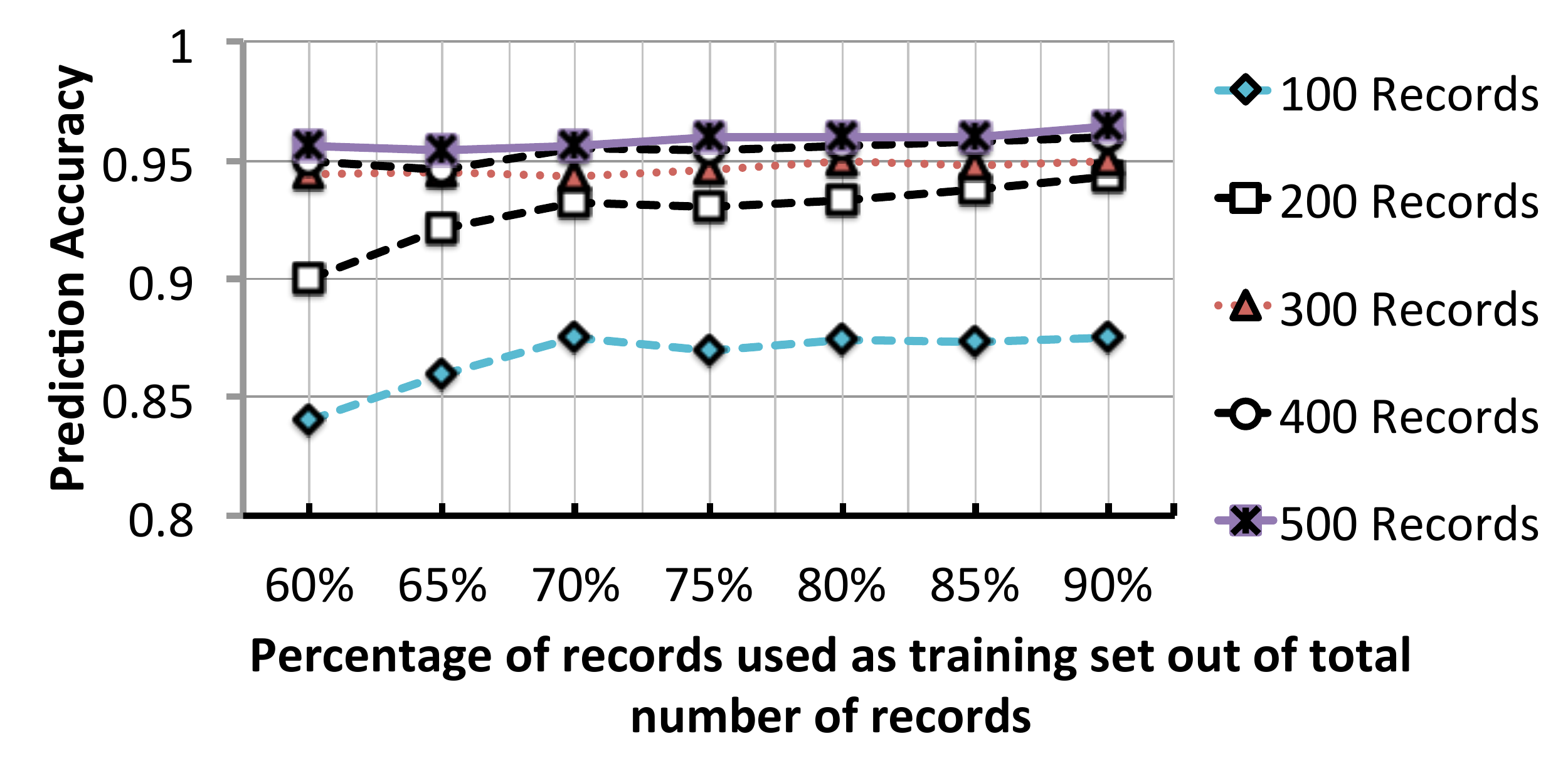}
    \caption{Prediction based on random dataset}
\end{figure}
Figure 10 illustrates the results of our evaluation using the parameter setting in Table 1. The x-axis shows the percentage records we have used as training set to predict the rest of the records. For example, the very first value on the bottom left of the graph shows the accuracy result based on a total of 100 query records, of which 60\%  records were used as the training set to predict the rest of the records (40\%). The prediction accuracy for this is around 84\%, as shown on the y-axis.

\subsubsection{Evaluating the Scheme on epSICAR Dataset}
We have also tested our prediction scheme using a subset of epSICAR dataset. We used 200 sequence records from the dataset. Each record consists of two context attributes: location and action. Each record is also associated with corresponding object (e.g., Hi-Fi Music system for listening music in living room), which can be considered as a service. The test result is shown in Figure 11. In the figure, when 30\% of the records (i.e., 60 records) were used as training set to predict the rest of the records, the accuracy of prediction was close to 100\% rate. 
\begin{figure}[!h]
  \centering
    \includegraphics[width=0.55\textwidth]{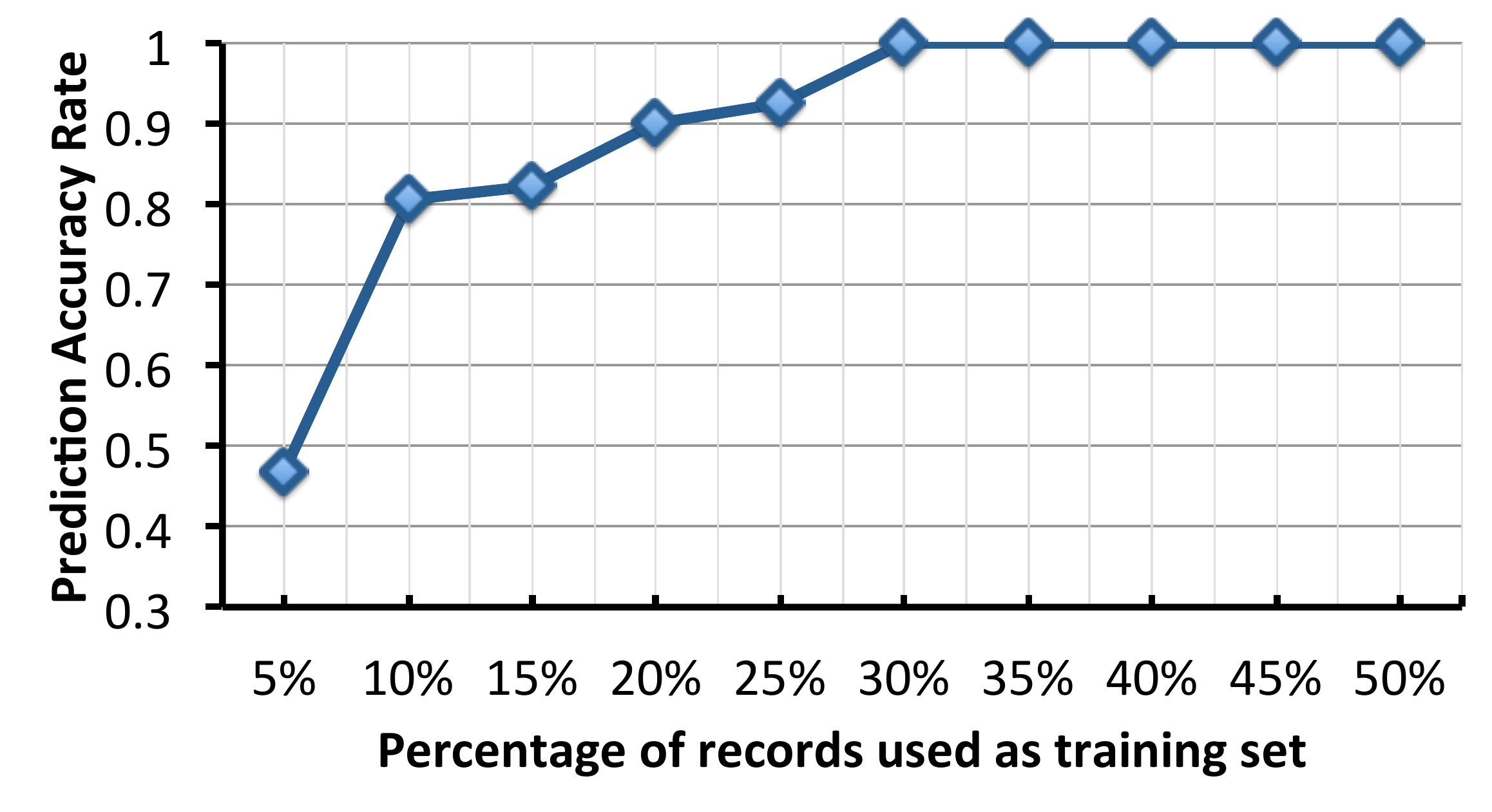}
    \caption{Prediction based on real dataset}
\end{figure}


\subsection{Trustworthy Service Discovery in MSNP}
This section presents the evaluation of the main component of the lightweight trustworthy service discovery scheme. The evaluation consisted of two parts:
\begin{enumerate}
\item	We evaluated the proposed scheme described in Section 6.2, in which a requester intends to obtain the trust score of a provider based on the requester's friends and friend of a friend (FOAF) (i.e., recommendation based on friends and FOAF). 
\item We evaluated the proposed scheme described in Section 6.3, in which a requester intends to obtain the trust score of a provider based on public proximal MSNP users who are non-friends of the requester (i.e. recommendation according to the public).
\end{enumerate}

We describe our evaluation approach below:
\begin{enumerate}
\renewcommand{\labelenumi}{(\theenumi)}
\item For each user record of a trust rating dataset, we considered the user as a \textit{requester} in MSNP who had a set of trust rating records (denoted by \textit{R-set}) which corresponds to the Reputation Rating Data ($RD$) in Definition 10.
\item From the \textit{R-set}, we separated the records into two subsets: \textit{rating of friends} and \textit{rating of non-friends}.
\item From the \textit{rating of non-friends} subset, we used the proposed schemes (described in Section 6.2 and 6.3) to predict what was the \textit{requester}'s rating for each \textit{rating of non-friends}. 
\item We also used the basic schemes (i.e., by simply referring to  the ratings from all the \textit{rating of friends} or all the friends of the corresponding users of \textit{rating of friends}) to predict what was the \textit{requester}'s rating for each \textit{rating of non-friends}. Then we compared the results between the proposed schemes with the basic schemes.
\item Finally, we compared the data transaction costs between the proposed schemes with the basic schemes. We then applied a basic CPI model to compare the schemes.
\end{enumerate}

In order to evaluate the proposed trustworthy service discovery scheme for MSNP, we have used the Advogato\footnote{\url{http://www.trustlet.org/wiki/Advogato_dataset}} dataset to simulate a large number of MSNP users' trust rating data. 

Advogato dataset is part of the Trustlet project  \cite{Massa2009}, which collects the trust rating values of social network site users since October 13 2007. Each record in the Advogato dataset consists of:
\begin{itemize}
\item	The ID of the person who rated another person
\item	The ID of the person who has been rated
\item	The rating value, which has three possible levels suggested by \cite{Massa2009}: Apprentice (represented by a score of 0.6), Journeyer (represented by a score of 0.8), and Master (represented by a score of 1.0).
\end{itemize}

We have tested our proposed trustworthy service discovery scheme using the Advogato dataset of 26 May, 2013. The original dataset contains many records with empty rating values (Some users have not rated any other users). Since our proposed scheme requires a fair number of rating data to calculate the trust score of a person based on other users' ratings, we have removed users who have less than 10 rating records from the original dataset.

The original Advogato dataset does not specify the relationship between users (i.e., are they real friends or not?). However, from their trust ratings, we categoried the relationship of users into two groups: when two users rated each other as `Master' level, they are `friends'. Otherwise, they are `non-friends'.  

The following sections present the evaluation cases and results.

\subsubsection{Selecting Recommender Based on Friends and FOAF}

The aim of this test is to show that the proposed schemes (described in Section  6.2) require less transaction cost but still can provide similar trust score measurement result as the basic schemes. 

The \textit{basic schemes} use a simpler approach to determine a service/content provider's trustworthiness based on the reputation rating of all the requester's friends or all the requester's FOAF. They are:
\begin{itemize}
\item \textbf{All Friends (AF)}.\\In this scheme, the requester computes a service provider's trust score based on the average rating values of all the requester's friends who have rated the service provider. 
\item \textbf{All Friends of Friends as Recommended Reference (AFOAF)}\\In this scheme, the requester computes a service provider's trust score based on the average rating value of all Recommended References (RR) of the requester's friends. The RR in this scheme are simply the FOAF who have rated the service provider without additional filtering. 
\end{itemize}
The \textit{proposed schemes}, which correspond to Step 2, 3 and 4 of \textbf{Algorithm 1} are:
\begin{itemize}
\item \textbf{One High Experience Friend (HEF)}\\In this scheme, the requester computes the service provider's trust score based on one single High Experienced Friend found from the requester's friends who have rated the service provider, and have largest rating records in the friends. HEF corresponds to the description in  \textbf{Algorithm 1}, Step 2. 
\item \textbf{One High Experienced FOAF (HEFHEF)}\\In this scheme, the requester computes the service provider's trust score based on one single High Experienced FOAF who has rated the service provider. The High Experienced FOAF is a friend of a HEF who may not have rated to the service provider, but the HEF has one or more friends who have rated the service provider. This scheme corresponds to \textbf{Algorithm 1}, Step 4. 
\item \textbf{One Most Similar Friend (MSF)}\\In this scheme, the requester computes service provider's trust score based on one single most similar friend. This scheme corresponds to \textbf{Algorithm 1}, Step 3. 
\end{itemize}

In this test case, we firstly retrieved a list of user IDs (as requesters) from the dataset. Each user had a list of ratings consisting of the IDs of the persons who had been rated, and the corresponding rating level value. Our test focused on predicting the requester's rating of each `non-friends' (representing service providers who will be evaluated by the requester) based on `friends' and `FOAF'.  

We used the above five different schemes to perform the prediction to show that the proposed scheme, which utilises the High Experience Friend's rating and the High Experience Friend's Recommended Reference person's rating (Algorithm 1) are efficient approach to measure the trust score of a provider. 

We assumed that the requester has replicated friends' $RD$ in local memory previously. Hence, at runtime, it can identify recommenders for computing the reputation score of a service provider without retrieving all friends' $RD$ directly from the friends' MWS or their cloud storages. The replicated $RD$ can only be utilised to identify recommenders. In order to find out the up-to-date reputation rating score from the recommenders, the requester still has to perform the request to retrieve the necessary $RD$ directly from the friends' MWS or their cloud storages. Depending on the scheme used, the required $RD$-retrieval process can be different. 

Table 2 summaries the cases of different schemes that were used for testing and comparison. The Comparable Count in the table represents the total number of rating records that have been used to test the scheme. Because each scheme relies on different criteria, the Comparable Count differs. For example, not all the users have available friends or FOAF's ratings to predict the trust rating of a specific user. Hence, such incomparable records have been excluded in the testing for that scheme.   

\begin{table}[!h]
\centering
  \begin{tabular}{ | >{\small}p{1.2cm} | >{\small}p{2.6cm} | >{\small}p{1.7cm} | >{\RaggedRight\small}p{3cm} |}
    \hline
     \textbf{Scheme} & 
     \textbf{Comparable Count} & 
     \textbf{Prediction Accuracy} & 
     \textbf{Average Minimum Transaction Required}
    \\ \hline \hline
    \multicolumn{4}{|l|}{\textit{Basic}} 
    \\ \hline
    AF & 1010 & 0.633569 & 6 
    \\ \hline
    AFOAF & 1075	& 0.642984	& 36 
    \\ \hline
    \multicolumn{4}{|l|}{\textit{Proposed}} 
    \\ \hline
    HEF &1010&0.635335&1
    \\ \hline
    HEFHEF &1010&0.640418& 6
    \\ \hline
    MSF &1010&0.579199&1
    \\ \hline
  \end{tabular}
  \caption{Comparison of Trust Schemes' Accuracy and Transaction Costs of Friends and FOAF}
\end{table}
The values of `Average Minimum Transaction Required' in Table 2 were computed as follows:
\begin{itemize}
\item \textbf{AF} scheme requires up-to-date reputation rating values from all friends. The \textit{average minimum transaction required} is equal to the average number of friends of  each requester under test, in which the average number of friends each requester has is 6, which is the average number of `Master' level ratings of each user in the Advogato dataset.

\item \textbf{AFOAF} scheme requires the highest transaction cost at runtime incurred by retrieving the up-to-date reputation rating values from all FOAFs. The total cost of the required transaction was the number of friends multiplied by the number of FOAF, which is 36.

\item In \textbf{HEF} scheme, since the requester has replicated the $RD$  previously, the replicated old $RD$ is sufficient for the requester to identify a HEF at runtime without consuming data transaction cost on retrieving new $RD$ via the Internet. Once a HEF is found, the requester only needs to retrieve the up-to-date reputation rating value from the HEF. Hence, in this case, the transaction cost is 1.

\item \textbf{HEFHEF} scheme requires the minimum transaction  values is 6, which is the sum of the transaction cost of retrieving RD from all friends of HEF.

\item \textbf{MSF} scheme incurs the same transaction cost as the HEF-based scheme.
\end{itemize}

\begin{figure}[!h]
  \centering
    \includegraphics[width=0.7\textwidth]{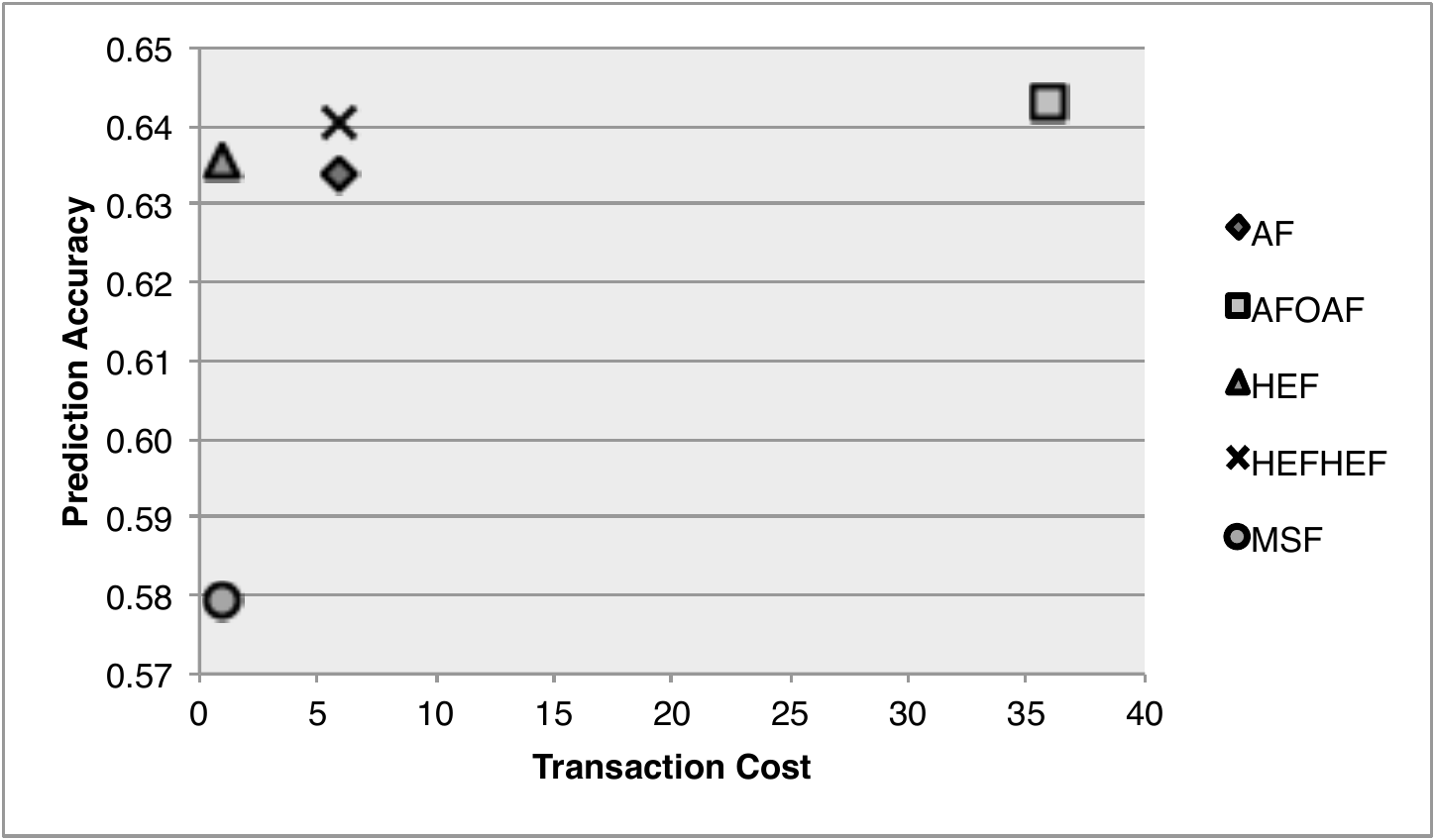}
    \caption{Predictive Rating Accuracy Comparison of Different Schemes based on Friends and FOAF}
\end{figure}

Figure 12 summarises and compares the prediction accuracy and the transaction cost of the five schemes in graphical form.

\begin{figure}[!h]
  \centering
    \includegraphics[width=0.7\textwidth]{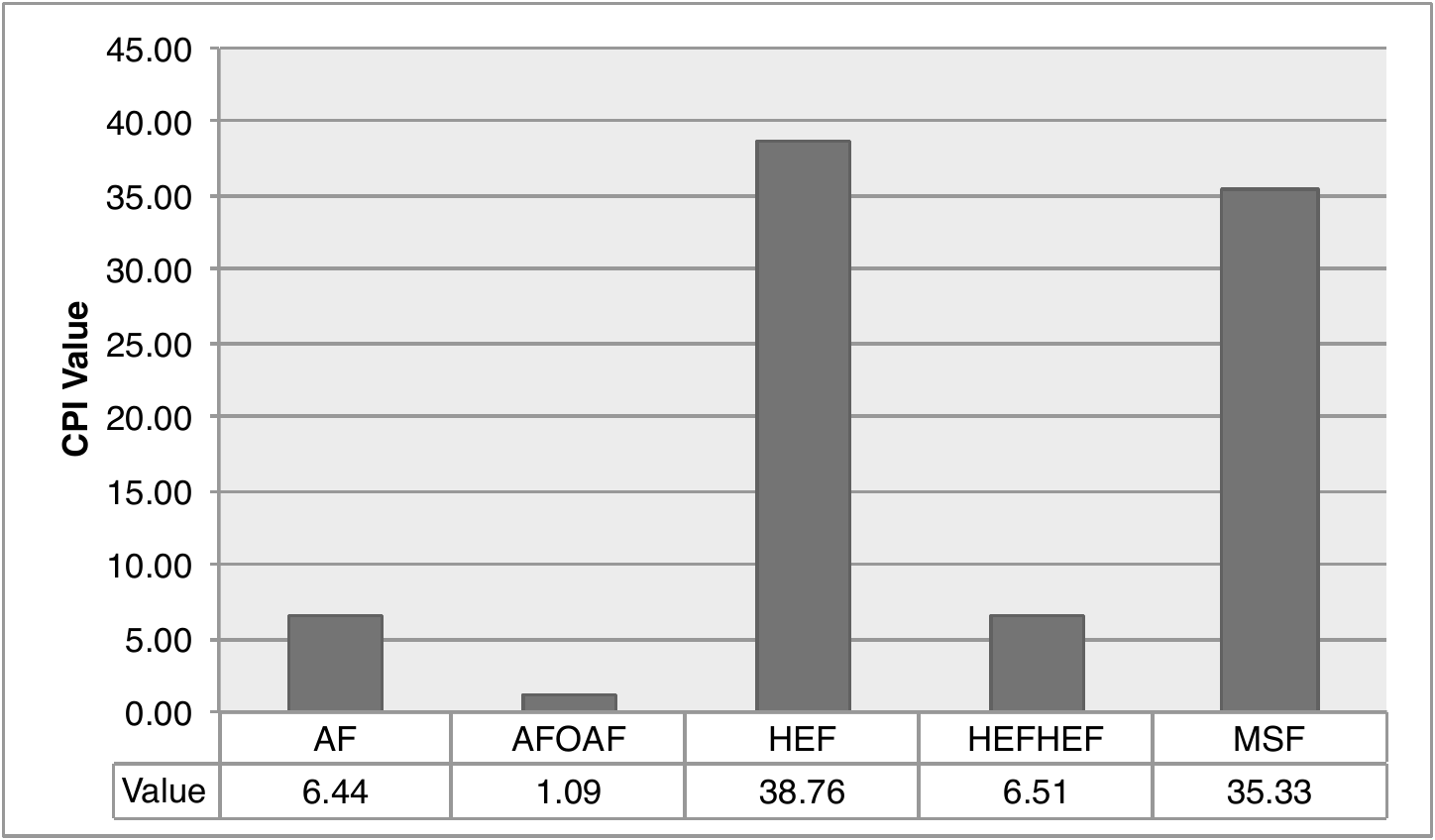}
    \caption{Cost and Performance Comparison of Different Schemes  based on Friends and FOAF}
\end{figure}

In order to highlight the overall improvement of the proposed approaches (HEF, HEFHEF, MSF) compared to the basic approaches (AF, AFOAF), we have translated the results into a cost and performance index (CPI) model. Figure 13 shows the cost-performance index value of each approach. As the figure shows, when direct friends are available as the recommenders of the reputation rating, the proposed HEF and MSF schemes provide better CPI values than the basic scheme---AF. When direct friends cannot be the recommenders, during which FOAF is needed, the proposed HEFHEF scheme gives a better CPI value than the general AFOAF scheme. 

\subsubsection{Selecting Recommenders Based on the Public}

The test described in this section corresponds to the scheme described in \textbf{Algorithm 2}, in which a requester is unable to determine a service/content provider's trustworthiness based on friends or FOAF's reputation rating values. Hence, the requester will refer to proximal strangers for the reputation rating values. However, the reputation rating of random selected stranger is unreliable. Hence, we presented in Section 6.3 an approach to identify which strangers' reputation rating values are reliable based on the stranger's \textit{experiences} and \textit{credibilities}.

This test aims to show that the proposed scheme can improve the accuracy when the trustworthy service discovery process is based on public proximal MSNP participants' rating scores. Recall that in our setting, each user in the Advogato dataset has `friends' (people who have been rated as `Master' level) and `non-friends' (people who have been rated as `Apprentice' or `Journeyer' level). In this test case,  we used the `non-friends' as the proximal strangers of the requester.

The test case compared the proposed scheme with the basic Na\"{i}ve scheme. The two schemes are summarised below:

\begin{itemize}
\item \textbf{Na\"{i}ve Scheme}\\The requester computes a service provider's trust score based on the average rating values of all the requester's `non-friends' who have rated the service provider. The service provider is excluded from the list of `non-friends'.
\item \textbf{Proposed Scheme}\\The requester computes a service provider's trust score based on a selected recommender based on both credibility and experience computed from the `non-friends' list. Same as the Na\"{i}ve scheme, the service provider is excluded from the list of `non-friends'.
\end{itemize}

We also included two additional schemes---Experience Only (Exp Only) and Credibility Only (Credit Only)---in which the requester selects a recommender based on only experience and based on only credibility respectively. These two schemes were included because we wish to show that the proposed scheme (based on both credibility and experience) provides better prediction accuracy than the cases of only using one of them to predict the reputation rating value.

When referring to the ratings from the public, the average minimum transactions required were the same, because the requester had to collect all the proximal MSNP participants' rating data in order to identify their credibility and experience. The value---7 is the average number of `non-friends' that each user had in the Advogato dataset.

In our test, we removed all the friends from the dataset. Each requester derived another user's rating score based on other user's rating values (i.e., public recommendations).

\begin{center}
\begin{table}[!h]
\centering
  \begin{tabular}{ | >{\small\raggedright}p{3cm} | >{\small}p{1.6cm} | >{\small}p{1.5cm} | >{\small}p{3cm} |}
    \hline
     \textbf{Scheme} & \textbf{Comparable Count} & \textbf{Prediction Accuracy} & \textbf{Average Minimum Transaction Required} 
    \\ \hline \hline
    Proposed Scheme using both Credibility and Experience & 851 & 0.703078 & 7 
    \\ \hline
    Na\"{i}ve Scheme &851&0.504942&7
    \\ \hline
   Experience Only Scheme&851&0.686321&7
    \\ \hline
    Credibility Only Scheme &851&0.499681&7
    \\ \hline
  \end{tabular}
  \caption{Comparison of Trust Schemes' Accuracy and Transaction Costs of Public}
\end{table}
\end{center}

\begin{figure}[!h]
  \centering
    \includegraphics[width=0.7\textwidth]{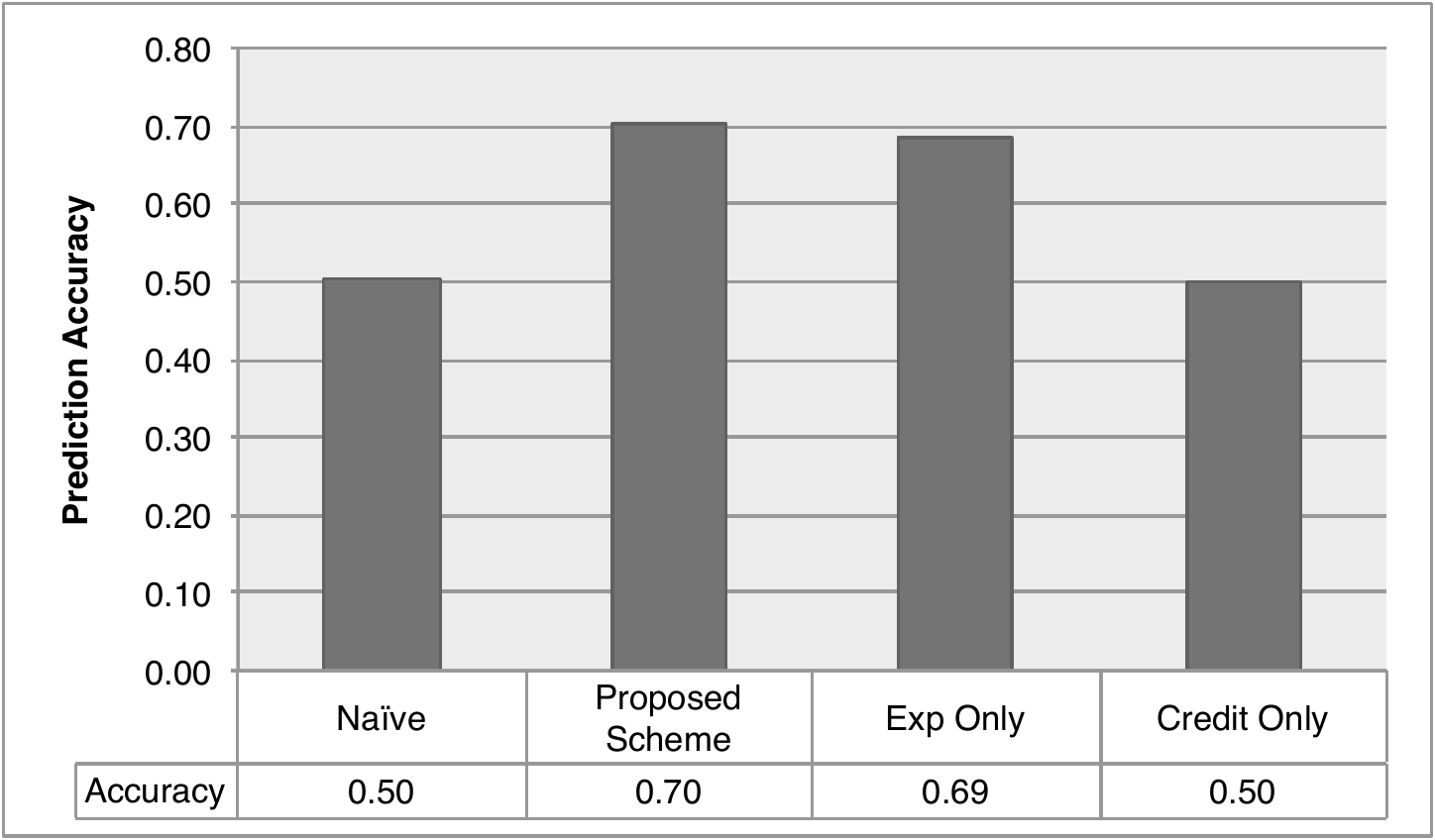}
    \caption{Predictive Rating Accuracy Comparison  of Different Schemes  based on Public}
\end{figure}

Table 3 shows the tabulated results, and Figure 14 the results in graphical form.

Since the transaction cost of all schemes were the same, we did  not need to calculate their cost-performance index value to compare their performance in this case. As the result shows, the accuracy of the Na\"{i}ve scheme was 50\%, which means that if the requester computes a provider's trust based on the average trust rating scores from all the proximal MSNP participants, it will only have a 50\% chance for the result to match what the requester expects. If the requester computes the provider's trust score based on the most experienced MSNP participant's rating (Exp Only), there is a 69\% chance that the result will match what the requester expects. On the other hand, if the requester only refers to the trust score of the highest credible MSNP participants (Credit Only), there is only a 50\% chance that the result can match what the requester expects. Our proposed scheme which combines experience with credibility outperforms the other schemes with a 70\% chance. Overall, all these schemes perform better than the Na\"{i}ve scheme in terms of accuracy.

The proposed scheme is shown to improve the accuracy when the prediction is based on public proximal MSNP participants' rating scores. However, since the rating score was computed based on strangers' ratings, the scheme was unable to reduce the transaction cost like the schemes based on friends and FOAF did. Because the requester did not have strangers' ratings pre-stored in its local memory or its cloud storage, in order to identify and compare the experience of all the proximal MSNP participants, the requester had to collect all the rating data from all the proximal participants' $agents$. Reducing the transaction cost in public-based trustworthy service discovery for MSNP requires further investigation. We consider this as one of our future research directions.